\documentclass[]{aa} 

\usepackage{graphicx}
\usepackage{epsfig}
\usepackage{natbib}
\usepackage{url}
\usepackage{amssymb}
\newcommand{\kms}{{\mathrm{km~s^{-1}}}}
\newcommand{\Teff}{{\mathrm{T_{eff}}}}
\newcommand{\FeH}{{\mathrm{[Fe/H]}}}
\newcommand{\logg}{\log g}
\newcommand{\alphaFe}{{[\alpha/\mathrm{Fe]}}}
\newcommand{\Vr}{{V_\mathrm{R}}}
\newcommand{\Vt}{{v_\mathrm{t}}}
\newcommand{\Vgc}{{V_\mathrm{GC}}}

\begin{document}

\title{Metallicity and kinematics of the bar in-situ
  \thanks{Based on ESO-VLT observations 079.B-0264, 060.A-9800 and 083.B-0767.}
  \thanks{Full Table \ref{tab:obj} is available in electronic form at the CDS via anonymous ftp to cdsarc.u-strasbg.fr (130.79.128.5) or via http://cdsarc.u-strasbg.fr/viz-bin/qcat?J/A+A/}
}

\author{C. Babusiaux\inst{1} 
	\and
	D. Katz\inst{1} 
	\and 
	V. Hill\inst{2}
	\and
	F. Royer\inst{1}
	\and 
	A. G\'omez\inst{1}
        \and 
	F. Arenou\inst{1}
	\and 
	F. Combes\inst{3}
        \and 
	P. Di Matteo\inst{1}
	\and	
	G. Gilmore\inst{4}
	\and
	M. Haywood\inst{1}
	\and
	A.C. Robin\inst{5}
	\and	
	N. Rodriguez-Fernandez\inst{6}
	\and
	P. Sartoretti\inst{1}
	\and 
	M. Schultheis\inst{2,5}
}
\institute{GEPI, Observatoire de Paris, CNRS, Universit\'e Paris Diderot ; 5 Place Jules Janssen 92190 Meudon, France
              \email{Carine.Babusiaux@obspm.fr}
	\and
Laboratoire Lagrange, University of Nice Sophia Antipolis, CNRS, Observatoire de la  C\^ote d'Azur, B.P. 4229, 06304 Nice Cedex 4, France
	\and
LERMA, Observatoire de Paris, CNRS, Universit\'e Pierre et Marie Curie ; 61 Av. de l'Observatoire, 75014 Paris, France
	\and
Institute of Astronomy, University of Cambridge, Cambridge, CB30HA, UK
	\and
             Institut Utinam, CNRS UMR6213, OSU Theta de Franche-Comt\'e Bourgogne, Universit\'e de Franche-Comt\'e, BP1615, 25010 Besan\c{c}on, France 
        \and
             CESBIO (UMR 5126), Observatoire de Midi-Pyr\'en\'ees, (CNES, CNRS, Universit\'e Paul Sabatier, IRD), 18 Av Ed. Belin,  BP2801, 31401 Toulouse cedex 9
}

\date{Received ; accepted } 

\abstract
{Constraints on the Galactic bulge and bar structures and on their formation history from stellar kinematics and metallicities mainly come from relatively high-latitude fields ($\vert b \vert>4\degr$) where a complex mix of stellar population is seen. } 
{We aim here to constrain the formation history of the Galactic bar by studying the radial velocity and metallicity distributions of stars in-situ ($\vert b \vert \leq 1\degr$). }
{We observed red clump stars in four fields along the bar's major axis ($l=10\degr$,$-6\degr$,$6\degr$ and $b=0\degr$ plus a field at $l=0\degr$, $b=1\degr$) with low-resolution spectroscopy from FLAMES/GIRAFFE at the VLT, observing around the \ion{Ca}{ii} triplet. We developed robust methods for extracting radial velocity and metallicity estimates from these low signal-to-noise spectra. We derived distance probability distributions using Bayesian methods rigorously handling the extinction law. }
{We present radial velocities and metallicity distributions, as well as radial velocity trends with distance. We observe an increase in the radial velocity dispersion near the Galactic plane. We detect the streaming motion of the stars induced by the bar in fields at $l=\pm6\degr$, the highest velocity components of this bar stream being metal-rich ([Fe/H]$\sim0.2$~dex). Our data is consistent with a bar that is inclined at 26$\pm3\degr$ from the Sun-Galactic centre line. 
We observe a significant fraction of metal-poor stars, in particular in the field at $l=0\degr$, $b=1\degr$. 
We confirm the flattening of the metallicity gradient along the minor axis when getting closer to the plane, with a hint that it could actually be inverted.
}
{Our stellar kinematics corresponds to the expected behaviour of a bar issued from the secular evolution of the Galactic disc. The mix of several populations, seen further away from the plane, is also seen in the bar in-situ since our metallicity distributions highlight a different spatial distribution between metal-poor and metal-rich stars, the more metal-poor stars being more centrally concentrated. }

\keywords{Galaxy: bulge -- Galaxy: formation -- Galaxy: abundances -- Galaxy: kinematics and dynamics}

\maketitle

\section{Introduction}

Although it is now well established that the Milky Way is a barred galaxy, the precise structure of the Galactic bar is subject to debate, and its links with the other Galactic stellar populations are still largely unknown.
In the following, we call ``bulge'' the full structure that is present within the central regions ($\vert l \vert \lesssim 10\degr$) independently of its origin. This in fact includes several components. 

A single bar model with a given semi-major axis and position angle does not seem to reproduce all the observations at the same time \citep[e.g.][]{Robin12}. The dominant boxy shape of the bulge covers the central $\sim$2-3~kpc ($\vert l \vert \leq 10\degr$) with an angle with respect to the Sun-galactic centre between 20 and 30$\degr$ \citep[e.g.][]{WeggGerhard13, Cao13}, although a smaller angle has also been discussed (e.g. \citealt{Robin12}). It presents an X-shaped structure \citep{McWilliamZoccali10, Nataf10, Saito11}. The bar angle seems to flatten in the inner regions ($\vert l \vert \leq 4\degr$) \citep{Nishiyama05, Gonzalez11bar}. A nuclear bar is suggested in the central molecular zone ($\vert l \vert \leq 1.5\degr$) \citep{Alard01, Sawada04, RodriguezCombes08}. 
At $\vert l \vert \geq 10\degr$, a longer thinner bar has also been proposed with an angle of $\sim45\degr$ \citep{Hammersley00,Benjamin05,CabreraLavers08}.   

All these observations have been found to be reproduced well with a single complex structure with N-body simulations by \citet{MartinezValpuesta06}, \citet{MartinezValpuestaGerhard11} and \citet{GerhardMartinezValpuesta12}. In their simulations, a stellar bar evolved from the disc, and the boxy/peanut/X-shaped bulge developed from it through secular evolution and buckling
instability. They find that the long bar in fact corresponds to the leading ends of the bar in interaction with the adjacent spiral arm heads \citep[see also][]{RomeroGomez11}. The change in the slope of the model longitude
profiles in the inner few degrees is caused by a transition from highly elongated to more nearly axisymmetric
isodensity contours in the inner boxy bulge. Their derived nuclear star count map displays a longitudinal asymmetry that could correspond to the suggested nuclear bar.

The main structures of the inner Galaxy and its kinematics \citep{Kunder12,Ness13kine} are therefore now fairly well explained as being mainly shaped by secular evolution. In parallel the bulge has been known to present an old age \citep{Zoccali03,Clarkson08,Brown10} and to be enhanced in $\alpha$-elements \citep{McWilliam94, Zoccali06, Fulbright07, Lecureur07}, suggesting a short formation timescale, which at first seemed at odds with the secular formation scenario. 

A metallicity gradient is observed along the bulge's minor-axis at $\vert b\vert > 4\degr$ \citep{Frogel99, Zoccali08}, while no significant gradient in metallicity has been found in the inner regions \citep{Ramirez00, Rich07inner}. 
The far and near parts of the X-shaped bulge have been shown to share the same metallicity, but only bulge stars that are more metal-rich than [Fe/H]$>-0.5$~dex show the distance split (due to line of sight crossing the near and far sides of the peanut), implying that the disc from which the boxy bulge grew had relatively few stars with [Fe/H]$<-0.5$~dex \citep{Ness12,Uttenthaler12}.
\cite{Hill11} show that the metallicity distribution in Baade's window ($l=1\degr$, $b=-4\degr$) can be decomposed into two populations of roughly equal sizes: a metal-poor component centred on [Fe/H] = $-0.3$~dex with a large
dispersion and a narrow metal-rich component centred on [Fe/H] = +0.3 dex. 
This separation is also seen in the kinematics of Baade's window stars, the metal-rich population presenting a vertex deviation typical of bar-like kinematics, while the metal-poor population
shows more isotropic kinematics \citep{Soto07,Babusiaux10}. Along the bulge minor axis, the metal-poor population shows a constant velocity dispersion, while the metal-rich population goes from bar-like
high velocity dispersion to disc-like low-velocity dispersion when moving away from the Galactic plane \citep{Babusiaux10}. 
The metal-poor population shows an $\alpha$-enhancement that is quite similar to the thick disc \citep{Melendez08,AlvesBrito10,Hill11,Gonzalez11alpha,Bensby13}, while the metal-rich population shows a low [$\alpha$/Fe] similar to the thin disc. 
The difference is also seen in the ages of microlensed main-sequence or turnoff stars by \cite{Bensby13}: the metal-poor stars are old ($>$10~Gyr) while the metal-rich stars show a large spread in age. 

Therefore there seems to be a mix of populations in the bulge: the metal-rich population is associated to the X-shape and is fading when moving away from the plane, while these two populations are mixed close to the plane. In fact, even more populations than just two could be mixed within the Galactic bulge \citep{Ness13met,Bensby13}, and we actually expect all the stellar populations of the Milky Way to be mixed in the central regions (including the most metal-poor ones, e.g. \citealt{GarciaPerez13}), the proportions and mixing efficiency providing the imprint of the Milky Way formation history. 

With the aim of deriving the kinematical and chemical properties of the bar in-situ, we observed red clump stars in four fields located along the bar major axis, those fields previously studied with near-infrared photometry by \cite{Babusiaux05}. Section \ref{Sdata} presents the data (target selection and data reduction), Section \ref{Sspectra} describes the spectral analysis to derive radial velocities and metallicities, and Section \ref{Sdistest} presents the distance (and extinction) estimation. Section \ref{Sresults} describes the results and their analysis. A summary of our conclusions is provided in Section \ref{Sconclu}. 

\section{\label{Sdata}The data}

With GIRAFFE LR08 we observed Galactic bar red clump giant candidates from the CIRSI Galactic Bulge Survey \citep{Babusiaux05}.
Those fields are located along the major axis of the bar within the Galactic plane, at $l=+10\degr$, $+6\degr$, $-6\degr$, and $b=0\degr$. A bulge reference field at ($l=0\degr$, $b=1\degr$), presenting a very homogeneous extinction, is also observed and used as a comparison point. The CIRSI field at $l=-10\degr$ is too faint to be observed by GIRAFFE owing to the large distance of the stars combined with high extinction.
Considering that the field at $l=+10\degr$ could be near the end of the bar and the beginning of the pseudo-ring and that \cite{Nishiyama05} find variation in the asymmetric signature of the bar within $\vert l \vert \sim 4\degr$, we expect to have a sampling of the main bar major axis with our three bar fields. 

We made our observations with the LR08 set-up of FLAMES, which covers the wavelength range 820.6-940.0~nm with a resolution R=6500. It covers the \ion{Ca}{ii} triplet (8498.02, 8542.09, 8662.14 \AA), which allows radial velocities and metallicities to be derived with low signal to noise spectra. This wavelength range is used in several large surveys (RAVE, ARGOS, Gaia-RVS, Gaia-ESO Survey) and has been extensively studied. This set-up is also the reddest low-resolution set-up of GIRAFFE entering the near-infrared area, therefore reducing the impact of the extinction compared to optical wavelengths. 

\begin{table*}
\centering
\caption{List of the science exposures. Period gives the ESO observing period (COM corresponding to a commissioning run) $\mathrm{N_{OBs}}$ gives the number of exposures of 45 minutes each for each field. $I_{max}$ is the estimated target $I$ band maximum magnitude. The S/N range provided corresponds to the quartile distribution.}
\label{tab:ListObs}
\begin{tabular}{lrrlll}
\hline\hline
Field & Period & $\mathrm{N_{OBs}}$ & $\mathrm{N_{stars}}$ & $I_{max}$ & S/N \\
\hline
9P ($l=+9.58, b=-0.08$) & 83 & 6 & 105 & 18.5 & 7-18 \\
5PN ($l=+5.65, b=-0.27$) & 79+83 & 2+5 & 102 & 18 & 9-15 \\
C32 ($l=0.00, b=+1.02$) & 79 & 2 & 114 & 16.3 & 16-22 \\ 
5NN ($l=-5.75, b=-0.22$) & 79+COM & 9+2 & 108 & 18 & 7-16\\
\hline
\end{tabular}
\end{table*}

\subsection{\label{sec:target}Target selection}

\begin{figure*}
 \centering
 \includegraphics[width=18cm]{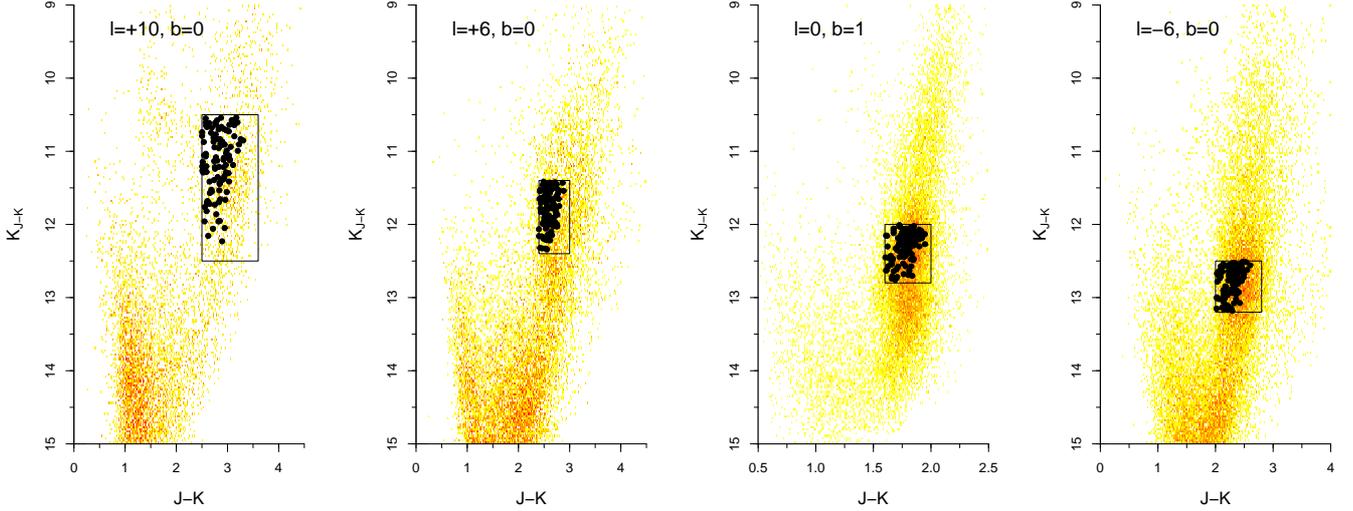}
 \caption{Near-infrared colour-magnitude diagrams of the fields. Sources observed by GIRAFFE are the filled dark circles.}
 \label{fig:CMD}
\end{figure*}

Stars were selected to be red clump candidates located in the bar according to the near-infrared photometry study of \cite{Babusiaux05}.
The selection box uses the reddening independent magnitude 
\begin{equation}
K_{J-K} = K - k_K/(k_J-k_K) (J-K),  
\end{equation}
with $k$ the extinction coefficients in the corresponding photometric band. Following \cite{Babusiaux05} for the target selection, we used $k_J=0.28$, $k_H=0.18$, and $k_K=0.11$ according to the extinction law of \cite{He95}.
The selected targets are represented in Fig. \ref{fig:CMD}.
Only stars with good near-infrared photometry were selected: goodness of fit $\chi^2<1.5$ and sharpness $\vert sharp \vert < 0.5$ in the three bands $J,H,K_s$. No other source has been detected within $1\farcs5$ of our targets.
Targets were also checked for having colours consistent with their being red clump stars. The $Q$ factor $Q = (J-H) - (k_J-k_H)/(k_J-k_K) (J-K)$ is independent of the extinction. Assuming $(J-K)_0 = 0.68$ and $(J-H)_0 = 0.61$ for a typical red clump star leads to an expected $Q_{RC}=0.184$. Stars with $\vert Q-Q_{RC}\vert<0.1$ were selected. 
To select stars with the best signal-to-noise ratio (S/N) possible, we selected stars with the lowest extinction, corresponding to the brightest estimated $I$ magnitude $I = K + (I-K)_0 + A_V (k_I-k_K)$, the $I$ band covering the wavelengths of our set-up. 
We give the maximum $I$ estimated magnitude of our sample for each field in Table \ref{tab:ListObs}. It has been computed with $(I-K)_0=1.4$ and $A_V$ estimated assuming the typical red clump star colours above. The number of exposures (OBs) of 45 minutes each are also provided in Table \ref{tab:ListObs}. They were computed to reach a S/N of 15 for the faintest stars of our selection. According to the S/N actually observed, we most probably have under-estimated the $I$-band extinction (see also Sect. \ref{Sdistres}).
All those selections were needed to avoid contamination of the sample by foreground red dwarfs and to ensure a minimum S/N in our spectra.

\subsection{\label{sec:targetseltest}Posteriori tests on the target selection}

\begin{figure}
 \centering
 \includegraphics[width=7cm]{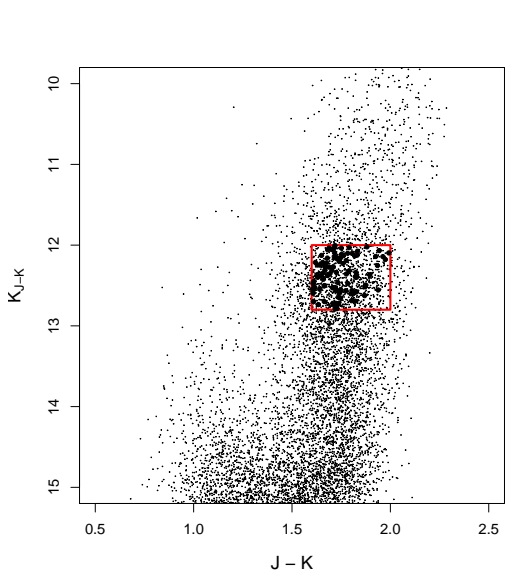}
 \caption{Simulated near-infrared colour magnitude diagrams of field ($l=0\degr$, $b=1\degr$) with the CMD selection box in red. This simulation was done using the \cite{Fux99} disc-particle distance distribution, a mean extinction $A_{550}=7.7$ with a dispersion of 0.8~mag, the isochrones of \cite{Bressan12}, a constant SFR, the \cite{Chabrier01} lognormal IMF, and Gaussian metallicity distribution centred at $-0.23$ with a dispersion of 0.5~dex. The number of stars is arbitrary. The field's black circles correspond to a hundred simulated stars selected using the same $Q$ and $I_{max}$ critera applied and detailed in section \ref{sec:target}.}
 \label{fig:SimulatedCMD}
\end{figure}

To study the impact of this complex target selection a posteriori on our final sample, we used a Monte-Carlo simulation on the field at ($l=0\degr$, $b=1\degr$). We used, as in Sect. \ref{Sdistest}, the isochrones of \cite{Bressan12}, a constant star formation history (SFR) and the \cite{Chabrier01} lognormal IMF. We followed the distance distribution provided by the \cite{Fux99} disc particles, with the Sun at 8~kpc from the Galactic centre. We used a Gaussian distribution to model the extinction using the mean and dispersion from Table \ref{tab:dist} and with the extinction law described in Sect. \ref{Sdistest}. As a first test we used a Gaussian metallicity distribution with the parameters of Table \ref{tab:Met}. The resulting colour-magnitude diagram (CMD) is presented in Fig. \ref{fig:SimulatedCMD}. 

We checked that our CMD box is large enough so that no bias is seen in the selected stars' metallicity or distance distributions. We checked the latter by shifting the N-body model distance distribution by 0.6~kpc (according to our results in Table \ref{tab:dist}), without changing the CMD box, and the mode of the selected distance distribution corresponds to the mode of the input distance distribution.

A bias in the metallicity distribution starts to be significant when the centre of the metallicity distribution is lower than $\sim -0.6$ dex. But of course if we change the mean metallicity of the sample significantly, we also see that the CMD box is shifted compared to the CMD red clump track. Using the two-component Gaussian distribution presented in Table \ref{tab:Semmul}, we checked that the bias against the metal-poor population is small.
We note, however, that this red clump CMD box leads to a strong bias towards younger ages: in our input constant SFR, the median age is 6.5~Gyr, while in the selected stars the median age is 4.5~Gyr. If we assume an age-metallicity correlation, such a red clump CMD box would therefore lead to a bias towards young metal-rich stars. 

We checked that our selection on the $Q$ factor does not remove any synthetic stars from the CMD box selection, even if the assumed red clump colours and extinction law are significantly different, indicating that this selection, done to remove foreground red dwarfs, is large enough. 
The selection on the estimated $I$ magnitude biases the sample toward lower extinction and shorter distance, as expected. The final synthetically selected sample distance mode is 7.4~kpc (corresponding to a 0.6~kpc bias and equivalent to our observed value in Table \ref{tab:dist}) and the extinction mode $A_{550}$ (absorption at 550~nm, see Section \ref{Sdistest}) is 7~mag (0.7~mag bias). No bias in the metallicity distribution seems to be created; however, a change in the age distribution is observed, this time favouring the old ages. 

We also tested our target selection using real data in the Baade's window red clump sample of \cite{Hill11}. Our selection using the $Q$ factor only removed 4\% of the Baade's window sample spread at all metallicities. Our way of estimating the $I$~magnitude from the near-infrared colours would not have implied any bias neither in the metallicity distribution in the Baade's window sample. 

As a conclusion, we expect a bias in our selection toward the lower extinction and shorter distance implied by our estimated $I$ magnitude cut, a cut needed to ensure a minimum S/N in our spectra. We do not expect any strong bias in the metallicity distribution, except against the most metal-poor tail. However, we could expect a bias in the age distribution, which is not possible to quantify easily owing to its high dependency on the used isochrones and the real extinction behaviour.

%
%

\subsection{Data reduction}

The observations were taken at different periods, as indicated in Table \ref{tab:ListObs}. Between periods 79 (Summer 2007) and 83 (Summer 2009), the CCD was changed \citep{Melo08}, significantly increasing the efficiency of the LR08 set-up. Our observations were used for commissioning this new CCD (in May 2008), allowing the needed S/N to be completed for the field at $l=-6\degr$.

The data reduction was performed using the GIRAFFE pipeline kit (version 2.8.7). Particular care has been taken for the dark-current subtraction in order to remove the electronic glow defect, present on the CCD detector in Period 79.  
The one-dimensional spectra were extracted with the optimal extraction method implemented in the pipeline. 

The sky contamination was removed using the algorithm described in \citet{Battaglia08}. The removed sky spectra is kept and used to remove pixels affected by strong sky emission in the analysis. 

The spectra have not been combined, so each epoch spectra is used individually in the following. This allows a transparent change in the CCD response between our different periods of observations, a filtering of the wavelengths affected by strong sky emission following its relative epoch radial velocity, and an optimal handling of the noise properties of each observation. 

We indicated a rough estimate of the S/N of our spectra (unused in the analysis) in Table \ref{tab:ListObs} 
following the proposed VO standard DER\_SNR \citep{dersnr}. 

\begin{figure}
 \centering
 \includegraphics[width=8cm]{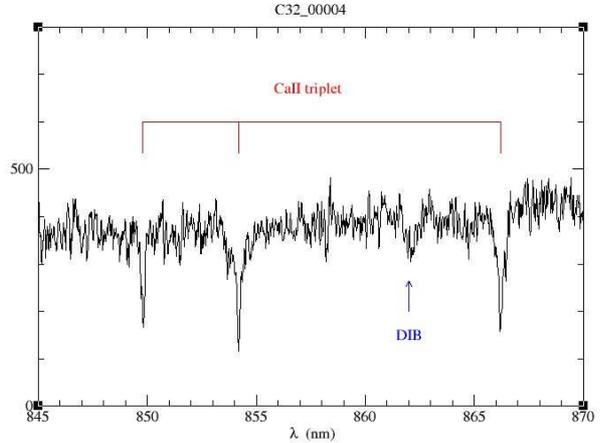}
 \caption{Example of a spectrum. This star is from the field at $l=0\degr$, $b=1\degr$, observed with a S/N of 21 and has a derived metallicity of [Fe/H]=$-0.6$~dex. The diffuse interstellar band (DIB) indicated is the strongest one at 862~nm.}
 \label{fig:speex}
\end{figure}

An example of a spectrum is presented in Fig. \ref{fig:speex}.

\section{\label{Sspectra}Spectral data analysis}

\subsection{\label{sec:spelib}Spectral library}

Two grids of synthetic spectra were generated and used for
two different purposes: (i)~derive the metallicities and
(ii)~assess the precision of the metallicities obtained.
The first grid is mono-dimensional, varying in metallicity
from [Fe/H]=$-1.5$ to $+1.0$~dex with a step of $0.1$~dex.
The atmospheric parameters are chosen as typical of the
red clump: $\Teff = 4750$~K, $\log g = 2.5$, and micro-turbulent velocity $\Vt=1.5~\kms$. Alpha-elements versus iron ratio
typical of the galactic trend is adopted: $\alphaFe=0.0$
for $\FeH \geq 0.0$, $\alphaFe = -0.4 \times \FeH$ for
$-1.0 \leq \FeH \leq 0.0$, and $\alphaFe = +0.4$~dex
for $\FeH \leq -1.0$~dex. 
Our spectra are dominated by the Ca alpha element through the \ion{Ca}{ii} triplet. Our adopted trend is consistent with the trend derived on bulge microlensed dwarfs by \cite{Bensby13} for this element.

The second grid is multi-dimensional but with a lower metallicity resolution. 
It extends in $\Teff$ from 4250 to 5250~K in steps of 250~K, $\log g = 2.0$ to 3.0
in steps of 0.5, $\FeH = -1.5$ to 0.5~dex in steps of 0.25~dex.
The micro-turbulent velocity and the $\alphaFe$
versus [Fe/H] relation are the same as in the first grid.
We checked with the synthetically selected stars presented in Fig. \ref{fig:SimulatedCMD} that 90\% of our stars should be within the grid ($4000<\Teff<5250$~K and $2<\log g <3$) and up to 98\% when considering the grid edge ($4500<\Teff<5500$~K and $1.5<\log g <3.5$).

The atomic parameters were taken from VALD \citep{Kupka00} and were checked on
Arcturus. Molecular lines of \element[][12]{C}\element[][14]{N} (private communication B. Plez 2010 and updated version with B. Edvardsson; 
the CN line list was assembled by B. Plez and was shortly described in 
\citealt{Hill02}), FeH \citep{Dulick03} and TiO (\citealt{Plez98}, considering 5 TiO isotopes from \element[][46]{Ti} to \element[][50]{Ti}) have been included in the computation of the synthetic spectra.

The model atmospheres were taken from the MARCS grid \citep{Gustafsson08}, computed in spherical geometry, and the synthetic spectra were computed with the Turbospectrum code \citep{Turbospectrum,TurbospectrumCode} in the wavelength range 835--895~nm.


%
%
%

Considering that we observed high-extinction fields, we also have strong diffuse interstellar bands (DIB) within our spectra \citep[e.g.][]{Munari08}. The velocity dispersion of the interstellar medium along the line of sight is too high to assume a single profile for the DIB fitting as done in \cite{HuiChen13}, and our spectra have too many lines and not enough S/N to try to extract the DIB equivalent width from stellar-spectra template subtraction. We therefore do not present here our attempts to extract the DIB information. However we identify the DIB wavelength ranges to remove those regions for the fits. For this we used the DIB profile generated following the prescription of \cite{Jenniskens94}\footnote{\url{http://leonid.arc.nasa.gov/DIBcatalog.html}} and removed wavelengths for which the DIB was absorbing more than 0.05\% of the flux. That leads to the following wavelengths being discarded : 8528.1--8533.7; 8613.9--8628.7; 8642.7--8654.1; 8761.8--8765.4~\AA. The two middle ones are the two strongest DIBs studied in \cite{Munari08}. The stronger DIB at 862~nm is indicated in Fig. \ref{fig:speex}.

\subsection{Radial velocities}


The radial velocities are derived by cross-correlation
with the synthetic spectrum typical of a solar metallicity
red-clump star ($\Teff = 4750$~K, $\log g = 2.5$), extracted from the library described in 
Sect.~\ref{sec:spelib}.

As summarized in Table~\ref{tab:ListObs}, between 2 and 11 exposures
were obtained per target. The spectra were not combined
prior to deriving the radial velocities. Instead,
a cross-correlation function (CCF) was derived for each
exposure and the CCF combined using the \cite{Zucker03}
maximum likelihood technique.

To optimize the processing time while preserving the numerical
precision, the CCF were calculated in two passes, by first shifting
the synthetic spectrum from $-$500 to $+$500~$\kms$ with a
step of 10~$\kms$. A second pass is then done within
$\pm$15~$\kms$ of the maximum of the first combined CCF with
a step of 1~$\kms$.
 
For each $\Vr$ step, spectrum normalization was done by fitting a cubic smoothing spline with four degrees of freedom on the spectra divided by the $\Vr$-shifted synthetic template. Wavelengths affected by strong sky emission (using an empirically tested quantile of 0.93 on the sky level) and by the DIB were eliminated from the normalization fit. Wavelengths affected by strong stellar absorption lines were also removed (normalized synthetic spectra intensity lower than 0.95). Bad pixels were eliminated in two passes: after a first normalization, hot pixels about six times the median absolute deviation (MAD) were removed and a new normalization done; pixels above and below (needed against too strong sky subtraction) four times the MAD were removed for the final normalization. 

The CCF was only computed on the wavelengths not affected by strong sky emission, DIB or hot pixels. 
The maximum of the combined CCF was derived by a second-order polynomial fit, around 70 $\kms$ from the maximum for the first pass, and on the full 15 $\kms$ range for the second pass. 
The theoretical errors were computed according to the prescription of \cite{Zucker03}. 
Stars with a bad fit during the maximum determination 
or a maximum CCF value lower than 0.1 were eliminated. 

\begin{figure}
 \centering
 \includegraphics[width=8cm]{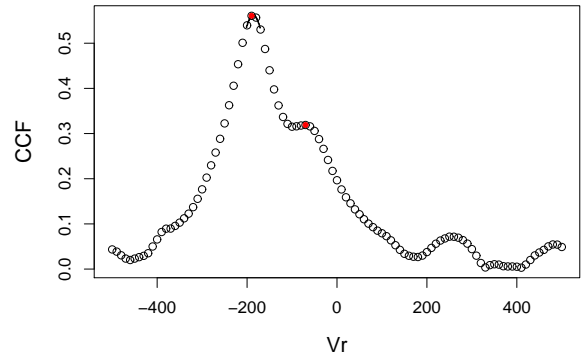}
 \caption{Combined cross-correlation function for C32\_00017. The filled red circles correspond to the local maxima of the CCF. This spectra is flagged as a blended one.} 
 \label{fig:ccf}
\end{figure}

Local maxima were detected in the first pass combined CCF (i.e. values higher than their 4 closest neighbours). When the value of CCF at those local maxima was higher than 50\% of the maximum of the CCF, they were flagged as blended spectra (see example in Fig. \ref{fig:ccf}).

The theoretical errors $\epsilon_{\Vr}$ were compared to the empirical ones $\epsilon(\Vr_{i})$ derived using the different exposures measurements $\Vr_i$. If the theoretical $\epsilon_{\Vr}$ were accurate, $(\Vr_{i} - \Vr) / \sqrt{\epsilon(\Vr_{i})^2+\epsilon_{\Vr}^2}$ should follow a unit normal distribution. Instead, the MAD of this distribution is 2.14$\pm$0.05 (value obtained using all the fields except C32, which only has two exposures; standard error obtained by bootstrap). No significant variation in this factor is seen as a function of S/N. The $\epsilon_{\Vr}$ used in the following of the paper were therefore the theoretical ones multiplied by 2.14. This error estimate does not take the template mismatch into account.

\subsection{\label{Smetest}Metallicities}

The S/N of our LR08 spectra range from
$\sim$7 to 22 (see Table~\ref{tab:ListObs}) and therefore do not contain enough
information to constrain all the atmospheric parameters
simultaneously. The analysis is therefore conducted in
two steps. First, the effective temperature, surface
gravity and micro-turbulence are assumed to be known from
the photometric selection process and to be typical of
the red clump ($\Teff = 4750$~K,
$\log g = 2.5$) and the metallicities are derived. 
This uses the first synthetic spectra grid presented in Sect. \ref{sec:spelib}.
In a second step, the
uncertainties on the metallicities due to possible
errors on the temperatures and gravity are assessed on the second 3D synthetic spectra grid.

The normalization is done in the same way as for the radial velocity determination. The only difference is that we remove only the strongest absorption lines (normalized synthetic spectra intensity lower than 0.8) to be able to deal with the highest metallicity templates.

As for the $\Vr$ determination, the observed and the synthetic spectra are compared on all pixels that have not been flagged as affected by strong sky emission, DIB, or cosmic rays. The cores of the \ion{Ca}{ii} lines are also eliminated. Indeed they are not good indicators of the atmospheric abundance of Ca and their modelling (NLTE, chromosphere, etc.) could suffer from larger uncertainty than the wings. The same boundaries as in \citet{Kordopatis11} were adopted for the cores of the \ion{Ca}{ii} lines: we eliminated 0.8, 1.2, and 1.2~\AA\ centred on the wavelengths 8498.02, 8542.09, and 8662.14~\AA, respectively. 

The metallicities are derived by minimum distance spectrum fitting, following TGMET \citep{Katz98} and ETOILE \citep{Katz01}. 
A quadratic sum of the residuals between the observed spectra $O$ and the synthetic spectra $S$ scaled by the continuum $C$ is computed on all the valid wavelengths $\lambda$. A weighted sum of those residuals is done on the different exposures $i$ using the median flux level $\mbox{med}_O$:
\begin{equation}
R = \sum_{i} {1 \over \mbox{med}_O} {1 \over n} \sum_{\lambda=1}^{n} \left({O(\lambda) - S(\lambda) C(\lambda)}\right)^2.
\end{equation}
The error variance variation with $\lambda$ can only be taken into account when a proper handling of the errors at each step of the calibration is done, see e.g. \cite{Koposov11}. 
Following \cite{Posbic12}, the minimum of the residuals is found by adjusting a third-degree polynomial due to the asymmetric profile of the residual as a function of metallicity around solar metallicity. 

\begin{figure}
 \centering
 \includegraphics[width=8cm]{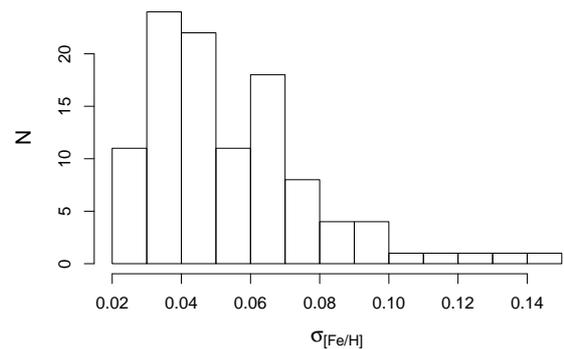}
 \caption{Distribution of the metallicity standard error as derived by bootstrap in field $l=0\degr$,$b=1\degr$. }
 \label{fig:metbootsig}
\end{figure}

Formal errors are derived by bootstrap. Their distribution in our reference field is presented in Fig. \ref{fig:metbootsig}.
We have checked our method using a synthetic spectra with added Gaussian noise down to a S/N of five for which we derived a bias free metallicity and bootstrap errors consistent with the Monte Carlo errors.  

\begin{figure}
 \centering
 \includegraphics[width=8cm]{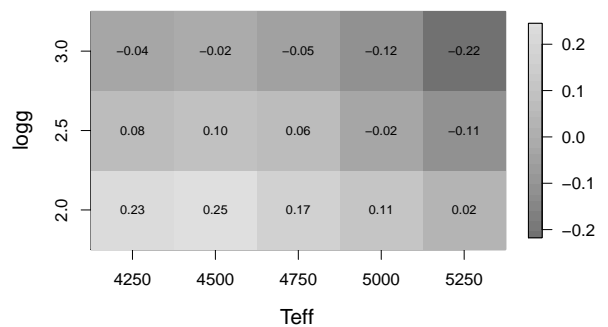}
 \caption{Metallicity derived for a synthetic spectra of S/N=50 (higher than our observed stars to test only the systematics) with [Fe/H]=0 and different $\Teff$ and $\logg$, using our reference grid done assuming $\Teff=4750$~K, $\logg=2.5$. }
 \label{fig:metSA}
\end{figure}

\begin{figure}
 \centering
 \includegraphics[width=7cm]{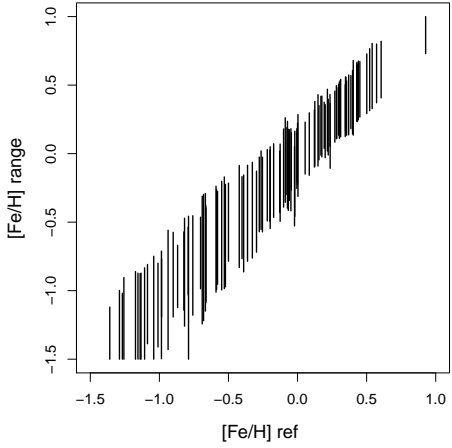}
 \caption{Sensitivity analysis of the derived metallicity when varying the assumed $\Teff$ and $\logg$ in field l=0$\degr$,b=1$\degr$. Each line shows the maximum and minimum metallicity derived against the reference metallicity obtained assuming $\Teff=4750$~K, $\logg=2.5$. }
 \label{fig:metSAC32}
\end{figure}

The metallicities have been derived up to now assuming fixed $\Teff=4750$~K and $\logg=2.5$.
We tested the sensitivity of our metallicity estimate on our choice of a fixed $\Teff$ and $\logg$ using the 3D synthetic spectral grid. 
Figure \ref{fig:metSA} shows the metallicity we derived with our method for a solar metallicity synthetic spectra for which we varied the $\Teff$ and $\logg$. 
Figure \ref{fig:metSAC32} shows the range of the metallicity estimate that we derived for each star of field $l=0\degr$,$b=1\degr$ using all the possible values of $\Teff$ and $\logg$ of our 3D grid. 

We also tested the sensitivity to the micro-turbulence by changing our default $\Vt=1.5$, value derived for Baade's window red clump by \cite{Hill11}, to $\Vt=2.0$~$\kms$,  which should cover the maximum plausible variation of $\Vt$ among red clump stars (\cite{deLaverny12} use $\Vt=2.0$~$\kms$ for all giants in their grid). The effect is similar to the measurement error: using $\Vt=2.0$~$\kms$ decreases all metallicities by 0.09~dex and adds a dispersion of 0.03~dex.

Overall, we conclude that a systematic error of about 0.2~dex should be added to our bootstrap (e.g. noise) error presented in Fig. \ref{fig:metbootsig}.

%
%

\section{\label{Sdistest}Estimating the distances and the extinction}

To estimate the distance distribution of our sample we used the classical Bayesian method \citep[e.g.][]{PontEyer04, JorgensenLindegren05, Burnett10, CBJ11}, 
particularly needed here where the S/N and wavelength coverage does not allow us to derive the atmospheric parameters of the individual stars.
We used the \cite{Bressan12} isochrones (Parsec 1.1) downloaded from CMD\footnote{\url{htpp://stev.oapd.inaf.it/cgi-bin/cmd}} version 2.5 with a step of 0.05 in logAge between [6.6, 10.13] and a step of 0.05 dex in [M/H] between [$-1.5$, 0.5] and 2MASS photometry. Each isochrone point $i$, corresponding to a metallicity [M/H]$_i$, age $\tau_i$ and mass $\mathcal{M}_i$, has a weight associated to it $P(i)$ according to the IMF $\xi(\mathcal{M})$ and SFR $\psi (\tau)$. We used here the \cite{Chabrier01} lognormal IMF (integrated over the mass interval between isochrone points) and a constant SFR (considering that we have a grid sampled in logAge this means that the SFR associated weight is proportional to the age), and we did not introduce any age-metallicity correlation:
\begin{equation}
P(i) \propto \int \xi(\mathcal{M}) d\mathcal{M} \int \psi (\tau) d\tau.
\end{equation}

We computed the probability of a star with the observed parameters $O$ ($\widetilde{[M/H]}$,$\tilde{J}$,$\tilde{H}$,$\tilde{K}$) to have the physical parameters of the isochrone point $i$ ($[M/H]_i$, $\tau_i$, $\mathcal{M}_i$, $\Teff_i$, $\log g_i$, $J_i^0$, $H_i^0$, $K_i^0$): 
\begin{equation}
P(i|O) \propto P(O|i) P(i). 
\end{equation}
To compute $P(O|i)$, we assume independent Gaussian ($\mathcal{N}$) observational errors on the metallicity and the magnitudes $\tilde{m}$. Assuming a distance $d$ and an extinction $A_{550}$ for the isochrone point $i$, we have
\begin{eqnarray}
P(O|i,d,A_{550}) & \propto & \mathcal{N}(\widetilde{[M/H]_i}-[M/H]_i,\epsilon_{[M/H]}) \nonumber \\
&& \prod \mathcal{N}(\widetilde{m}-m_i,\epsilon_m). 
\end{eqnarray}

\begin{figure}
 \centering
 \includegraphics[width=9cm]{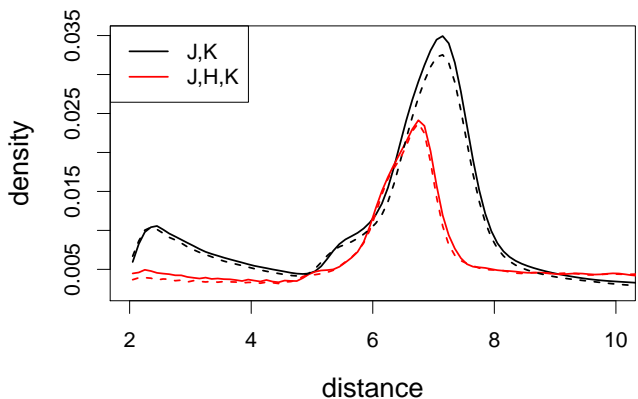}
 \caption{Distance probability distribution function for star C32\_00006 using only $J$ and $K$ magnitudes (black) or adding the $H$ band magnitude (red). In dotted lines we show the distribution obtained assuming the single reference metallicity (obtained by assuming $\Teff=4750$ and $\logg=2.5$). }
 \label{fig:distprobEx}
\end{figure}

Instead of using a single value $\widetilde{[M/H]}$ derived for a typical red clump star as in the previous section, we take advantage of the isochrone prior to help break the $\Teff$/$\logg$/[M/H] degeneracy illustrated in Fig. \ref{fig:metSAC32}. The $\widetilde{[M/H]}_i$ value is indeed the metallicity obtained for an isochrone point with ($\Teff_i$, $\logg_i$) by a 2D spline regression\footnote{R \{mda\} package} on the 3D synthetic spectra grid. But we see in Fig. \ref{fig:distprobEx} that this does not lead to significant improvement. 

Since the isochrone grid stops at [M/H]=0.5, we have to assume that the higher metallicity isochrones would be identical to the [M/H]=0.5 one and therefore add the probabilities corresponding to [M/H]$>$0.5 to the [M/H]=0.5 isochrone points. 

The apparent magnitude $m_i$ derived from the isochrone $i$ is a function of the absolute magnitude $M_i^0$, the extinction $A_m$, and the distance $d$:
\begin{equation}
m_i = M_i^0 + 5 \log d -5 + A_m.
\label{eq:Pogson}
\end{equation}
We therefore derived $P(O|i,d,A_{550})$ for a very thin 2-D grid of distances $d$ and extinction $A_{550}$. $A_{550}$ is the absorption at 550~nm, also often written $A_0$, and is roughly equivalent to $A_V$ \cite[e.g.][]{CBJ11}.
To derive the extinction in the different photometric bands $A_m$, we used the extinction law  $E_\lambda = 10^{-0.4 k_\lambda}$ of \cite{FitzpatrickMassa07}. We used a typical red clump SED $F_\lambda^0$ from \cite{CastelliKurucz03} ATLAS9 models, to be consistent with the isochrone colours. 
With $T_\lambda$ the photometric total instrumental transmission we have
\begin{eqnarray}
A_m &=& m - m^0 = -2.5 \log_{10}\left({F \over F^0}\right) \nonumber \\
   &=& -2.5 \log_{10}\left({\int F_\lambda T_\lambda E_\lambda^{A_{550}} d\lambda \over \int F_\lambda T_\lambda d\lambda}\right).
\end{eqnarray}
To take the non-linearity of the above equation into account, we used a discrete table of $A_m$ as a function of $A_{550}$. 
Considering neither the stellar SED nor the non-linearity of the relation between $A_m$ and $A_{550}$ can lead to 0.025 mag difference in distance modulus at $A_{550}=10$ mag. This is, of course, small since we are in the near-infrared, smaller than the uncertainties in the extinction law itself, but not fully negligible. 

As an example, with this computation we obtain $A_{550}$:$A_J$:$A_H$:$A_K$=1:0.237:0.141:0.086 and $A_K/E(J-K)$=0.567 for $A_{550}=7.7$ (corresponding to our mean value for field at $l=0\degr$,$b=1\degr$). This is closer to the empirical results obtained with red clump giants by \cite{Nishiyama08} than the values of the \cite{Cardelli89} extinction law calibrated on hot stars. 

What we seek is $P(d,A_{550}|O)$, which we obtain by marginalization over the isochrone points:
\begin{equation}
P(d,A_{550}|O) \propto \sum_i P(O|i,d,A_{550}) P(i).
\end{equation}

Marginalization over the extinction leads to $P(d|O)$. Figure \ref{fig:distprobEx} shows the resulting distance probability distribution function for a typical star of field $l=0\degr$,$b=1\degr$, with and without adding the $H$ band magnitude information. Although adding the $H$ band leads to a decrease in the probability of being a foreground star, we see that the solution is more degenerate. 
We indeed checked that the colour-colour relations (e.g. here $H-K$ versus $J-K$) around the Hipparcos red clump stars is not perfectly adjusted by the isochrone colours. It is therefore not surprising that adding the $H$ band information in fact increases the degeneracy rather than reducing it. We can also see that including $H$ band also changes the derived distance (by ~0.4~kpc). The exact distance indeed dependents on isochrone (and colour) but any resulting bias would be the same in all our fields. 

In the following we use the full distance probability distribution function. The typical confidence intervals at 1 $\sigma$ are between 0.6 and 0.8~kpc. However the blurring that this distribution will induce on the figures of sections \ref{Sdistres} and \ref{Svrdist} will be greater than this, owing to the degeneracies: only the peak of the distribution (corresponding to 50\% or 68\% of the probabilities) can be approximated by a Gaussian in most of the cases, but the distributions are often skewed and can present secondary peaks.

\section{\label{Sresults}Results and analysis}

The resulting catalogue is reported in Table \ref{tab:obj} available at the CDS. 

\begin{table*}[!ht]
 \centering
\caption{CIRSI photometry and derived heliocentric radial velocity and metallicity of our stars. The systematic 0.2~dex error on the metallicity has been taken into account in the provided $\epsilon_\mathrm{[Fe/H]}$. The full table is available in electronic form at the CDS.}
\label{tab:obj}
\setlength\tabcolsep{3pt}
\begin{tabular}{lccccccccrrrr}
\hline
\hline
Name & Right ascension & Declination &  $J$ & $\epsilon_J$ & $H$ & $\epsilon_H$ & $K$ & $\epsilon_K$ & $\Vr$ & $\epsilon_{\Vr}$  & [Fe/H] & $\epsilon_\mathrm{[Fe/H]}$ \\
     & J2000 & J2000 & mag & mag & mag & mag & mag & mag & \multicolumn{2}{r}{$\kms$~} & dex & dex \\
\hline
C32\_00004 & 17:41:16.298 & $-28$:29:26.851 & 14.827 & 0.017 & 13.502 & 0.017 & 13.153 & 0.051 & $-2.7$ & 1.2 & $-0.58$ & 0.21\\
C32\_00006 & 17:41:16.624 & $-28$:27:34.606 & 15.052 & 0.017 & 13.717 & 0.023 & 13.302 & 0.051 & $-232.7$ & 1.2 & $-0.20$ & 0.20\\
C32\_00012 & 17:41:17.318 & $-28$:30:32.357 & 15.086 & 0.018 & 13.814 & 0.020 & 13.482 & 0.051 & $-140.2$ & 1.7 & $-1.14$ & 0.21\\
C32\_00014 & 17:41:17.387 & $-28$:23:02.116 & 15.243 & 0.018 & 13.852 & 0.022 & 13.431 & 0.053 & $125.1$ & 1.0 & $0.93$ & 0.20\\
C32\_00018 & 17:41:17.561 & $-28$:30:04.326 & 15.128 & 0.016 & 13.773 & 0.017 & 13.310 & 0.051 & $64.1$ & 1.1 & $0.50$ & 0.20\\
C32\_00022 & 17:41:17.988 & $-28$:25:32.998 & 15.443 & 0.018 & 14.159 & 0.023 & 13.779 & 0.051 & $93.9$ & 1.7 & $-1.04$ & 0.21\\
C32\_00025 & 17:41:18.412 & $-28$:18:32.237 & 15.205 & 0.017 & 13.890 & 0.022 & 13.433 & 0.051 & $-7.2$ & 1.2 & $-0.00$ & 0.20\\
... & ... & ... & ... & ... & ... & ... & ... & ... & ... & ... & ... & ... \\
\hline
\end{tabular}
\end{table*}

\subsection{The radial velocities distribution}

\begin{figure}
 \centering
 \includegraphics[width=8cm]{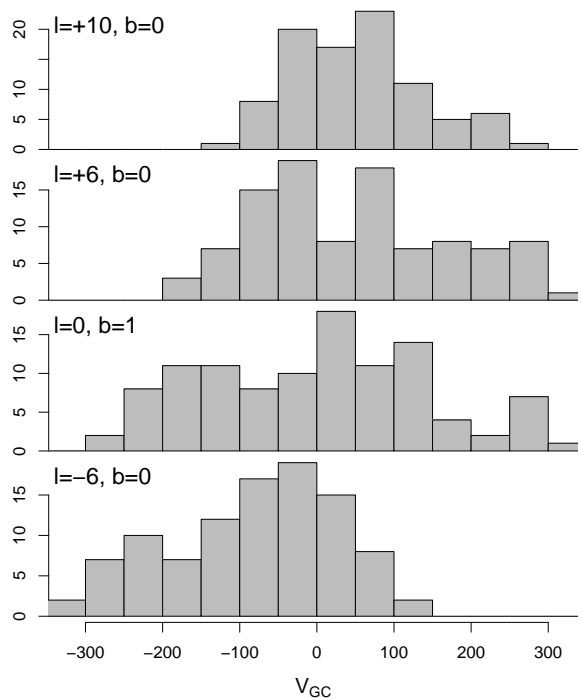}
 \caption{Histogram of the Galactocentric radial velocity distribution in our fields.}
 \label{fig:VrHist}
\end{figure}

Figure \ref{fig:VrHist} shows the radial velocity distribution of our fields for which Table \ref{tab:Vr} gives a summary. 
To correct for the solar reflex motion, we computed the Galactocentric velocity $\Vgc$ using the same formulae as  \cite{Beaulieu00}, \cite{Kunder12}, and \cite{Ness13kine}: 
\begin{eqnarray*}
 \Vgc &=& \Vr + 220 \sin l \cos b \\
 & & + 16.5 [\sin b \sin 25 + \cos b \cos 25 \cos(l-53)].
\end{eqnarray*}



\begin{figure}
 \centering
 \includegraphics[width=8cm]{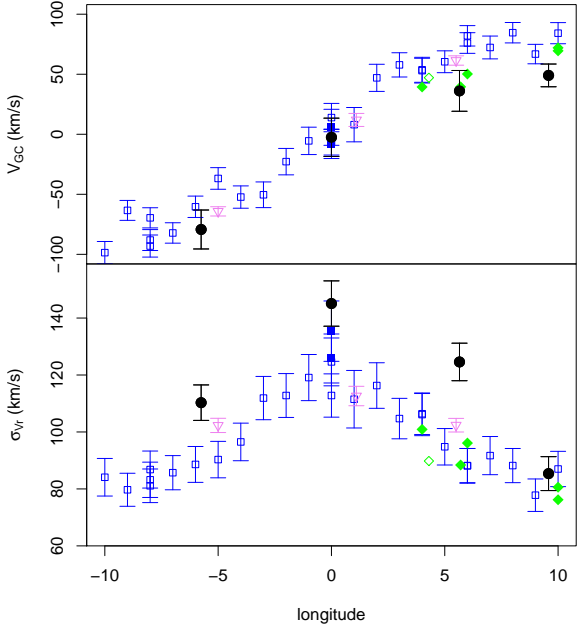}
 \caption{Mean galactocentric velocity (top) and velocity dispersion (bottom) of our fields (black filled circles), together with the results of \cite{Kunder12} (blue squares), \cite{Nidever12} (green diamonds), and \cite{Rangwala09a} (violet triangles). Filled symbols correspond to fields with $\vert b \vert \leq 2\degr$. }
 \label{fig:VrBrava}
\end{figure}

Figure \ref{fig:VrBrava} shows the comparison of the mean velocity and velocity dispersion of our fields with the BRAVA data \citep{Rich07brava, Kunder12}, the APOGEE first results of \cite{Nidever12}, and the data from 
\cite{Rangwala09a}. We distinguish in this figure the fields at $\vert b \vert \leq 2\degr$ and the fields at $b=-4\degr$. 
Apart from $l=10\degr$, our radial velocity dispersion is higher than the BRAVA data at $b=-4\degr$, consistent with dynamical models (e.g. \citealt{Fux99} and \citealt{Zhao96}) predicting an increase in the velocity dispersion closer to the Galactic plane, which can also be seen in the BRAVA data at $l=0$. However, our radial velocity dispersion is also higher than the results of \cite{Nidever12}, while the latter contains two fields ($l=4\degr$ and $l=6\degr$) at $b=0\degr$. This is most probably due to the different selection function and its consequence on the distance distribution (see Sect. \ref{Svrdist}).
For the field at $l=10\degr$ we see in Sect. \ref{Svrdist} that in fact a very large portion of our sample is contaminated by the end of the disc, explaining its low-velocity dispersion. 

\begin{table*}
\centering
\caption{Summary of the radial velocity distribution. $N$ gives the number of stars with good radial velocities (good CCF fit, no binary detection). $N_{blended}$ gives the number of stars removed from the sample due to multiple peaks in the $\Vr$ CCF. $\langle \Vr \rangle$ is the weighted mean heliocentric radial velocity and $\Vgc$ the galactocentric one. }
\label{tab:Vr}
\begin{tabular}{lrrllllll} 
\hline\hline
Field & $N$ & $N_{blended}$ & $\langle \Vr \rangle$ & $\langle \Vgc \rangle$ & $\sigma_{\Vr}$ & skew & kurtosis \\ 
 & & & $\kms$ & $\kms$ & $\kms$ & & & \\ 
\hline
$l=+10\degr, b=0\degr$ & 92 & 13 & 2 $\pm$  9 & 49 & 85  $\pm$  6 &  0.5  $\pm$  0.2 &  -0.3 $\pm$  0.3 \\ 
$l=+6\degr, b=0\degr$ & 101 & 1 & 4  $\pm$  17 & 36 & 125 $\pm$  7 &  0.4 $\pm$  0.1 & -0.9  $\pm$  0.2  \\ 
$l=0\degr, { }~b=1\degr$ & 107 & 6 & -12  $\pm$  15 & -3 & 145 $\pm$ 8 & 0.1 $\pm$  0.1 &  -0.8  $\pm$  0.2  \\ 
$l=-6\degr, b=0\degr$ & 99 & 9 & -65 $\pm$  15  & -79 & 110  $\pm$  6 &  -0.3 $\pm$  0.1 &  -0.9  $\pm$  0.2  \\ 
\hline
\end{tabular}
\end{table*}

\cite{Nidever12} detected a high-velocity component at around 200~$\kms$ in their fields, including $l=+6\degr$ and $l=+10\degr$. According to Fig. \ref{fig:VrHist}, this high-velocity component seems to be present in our fields with $l\geq0$. At $l=-6\degr$ there seems to be its counterpart: a high negative velocity component, as expected if indeed this component was associated with bar orbits as suggested by \cite{Nidever12}. We therefore applied the SEMMUL Gaussian-component decomposition algorithm  \citep{CeleuxDiebolt86} to our distributions as a clustering tool with two components and present the results for fields $l=\pm6\degr$ in Table \ref{tab:SemmulVr}.  At $l=10\degr$ forcing a two-component solution also gives a high-velocity component but with only 9$\pm3$\% of the sample, so not significant enough to be added in this table. At $l=0\degr$ we need a three-component solution to see a high-velocity component appearing but containing only 2\% of the sample. 
According to the Bayesian information criterion (BIC), the decomposition presented in Table \ref{tab:SemmulVr} is significant (better than a single component fit) only for field $l=+6\degr$. However if we remove the metal-poor stars with [Fe/H]$<-0.5$, this two-component decomposition also becomes significant for $l=-6\degr$. 
The two-component solution at $l=\pm6\degr$ gives a high-velocity component at $\vert \Vgc\vert  \sim 230~\kms$ with a velocity dispersion compatible with the results of \cite{Nidever12}. This high-velocity component has a mean metallicity of $\sim 0.2$~dex and represents about 20\% of our samples at $l=\pm6\degr$. The high-velocity component seems to be behind the main component at $l=+6\degr$ and in front of the main component at $l=-6\degr$.
This, combined with the correlation between distance and radial velocity study of Sect. \ref{Svrdist}, confirms the \cite{Nidever12} interpretation of this high-velocity component as being linked to the bar dynamics. 

\begin{table*}
\caption{SEMMUL two Gaussian-component decomposition on the radial velocity distribution. $\langle \Vgc \rangle$ is the mean galactocentric radial velocity and $\sigma_{Vr}$ the radial velocity dispersion. [Fe/H] is the mean metallicity of the components. The distance corresponds to the mode of the distance distribution function (in kpc).  $\sigma_\mathrm{dist}$ corresponds to the 68\% HDI of the distribution (see Sect. \ref{Sdistres}). The errors are computed using bootstrap resampling and do not include other local maxima in the decomposition.}
\label{tab:SemmulVr}
\begin{center}
\begin{tabular}{lllllll}
\hline\hline
Field & $\langle \Vgc \rangle$ & $\sigma_{\Vr}$ & \% & [Fe/H] & dist & $\sigma_\mathrm{dist}$ \\ 
 & $\kms$ & $\kms$ & & dex & kpc & kpc \\
\hline 
$l=+6\degr$ & -2 $\pm$ 18 & 84 $\pm$ 14 & 77 $\pm$ 9 & 0.12 $\pm$ 0.06 & 5.8 $\pm$ 0.1 & 2.0 $\pm$ 0.1 \\
 & 229 $\pm$ 22 & 47 $\pm$ 16  & 23 $\pm$ 9  & 0.23 $\pm$ 0.07 & 6.6 $\pm$ 0.1 & 1.9 $\pm$ 0.1\\
\hline 
$l=-6\degr$ & -235 $\pm$ 11  & 46 $\pm$ 9  & 23 $\pm$ 4 & 0.24 $\pm$ 0.10 & 9.2 $\pm$ 0.2 & 2.3 $\pm$ 0.1\\
   & -34 $\pm$ 18  & 75 $\pm$ 6  & 77 $\pm$ 4  & -0.06 $\pm$ 0.09 & 9.0$\pm$ 0.2 & 2.6 $\pm$ 0.2\\
\hline
\end{tabular}
\end{center}
\end{table*}

\subsection{Metallicity distribution function}

\begin{figure}
 \centering
 \includegraphics[width=8cm]{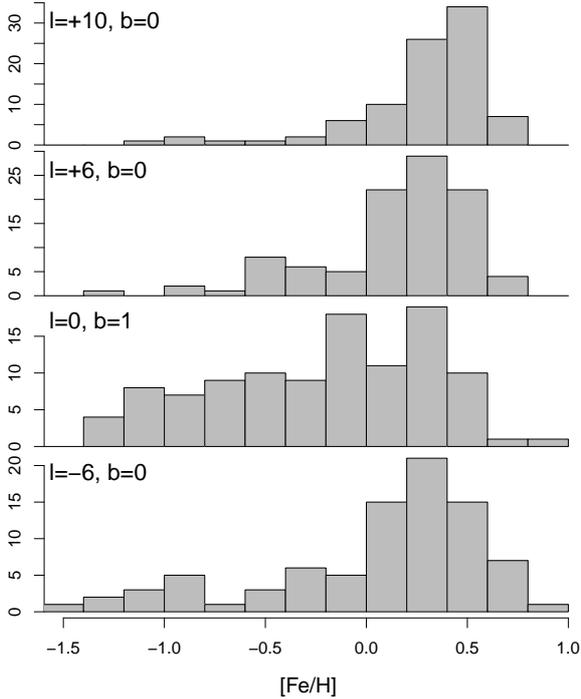}
 \caption{Histogram of the metallicity distribution in our fields.}
 \label{fig:MetHist}
\end{figure}

\begin{table}
\caption{Summary of the metallicity distribution.  $N$ gives the number of stars with derived metallicities. $\langle \mathrm{[Fe/H]} \rangle$ and $\sigma_{\mathrm{[Fe/H]}}$ are the mean and the dispersion of the metallicity distribution (in dex). }
\centering
\label{tab:Met}
\begin{tabular}{llll}
\hline\hline
Field & $N$ & $\langle \mathrm{[Fe/H]} \rangle$ & $\sigma_{\mathrm{[Fe/H]}}$ \\ 
\hline
$l=+10\degr, b=0\degr$ & 90 & { }~0.28 $\pm$ 0.04 & 0.34  $\pm$  0.05 \\ 
$l=+6\degr, b=0\degr$ &  100 & { }~0.15  $\pm$  0.04 & 0.37  $\pm$  0.04 \\ 
$l=0\degr, { }~b=1\degr$ & 107 & $-0.21$  $\pm$  0.05 & 0.54  $\pm$  0.03 \\ 
$l=-6\degr, b=0\degr$ & 88 & { }~0.03 $\pm$  0.06 & 0.55  $\pm$  0.05 \\ 
\hline
\end{tabular}
\end{table}

Figure \ref{fig:MetHist} shows the metallicity distribution of our fields for which Table \ref{tab:Met} gives a summary. 
All the metallicity distributions of our major axis fields can be decomposed into two Gaussian components, similar to what we found along the minor axis in \cite{Babusiaux10}. Table \ref{tab:Semmul} shows the results of the SEMMUL Gaussian-component decomposition algorithm \citep{CeleuxDiebolt86} applied to the [Fe/H] distribution of our fields. 
The size of our sample is, of course, too small, and the error on our metallicity estimates too large to conclude anything about the possibility of having more components, as proposed by \cite{Ness13met}. In particular we see that the metal-poor tail is included in the metal-poor component via an increase in the metallicity dispersion of this component, visible in particular at $l=+10\degr$. Moreover, the metallicities of the populations have no physical reason to be Gaussian. Such a decomposition should therefore simply be read as a clustering tool for studying the components' properties in the other dimensions (velocity, distance), but the mean values of those populations should be used with caution.

There is no significant difference in the metallicity distribution of the fields at $l=+6\degr$ and $l=-6\degr$ with a median metallicity of, respectively, $0.23\pm0.03$ and $0.20\pm0.04$~dex (Kolmogorov-Smirnov test p-value = 0.6). However, the field at $l=+10\degr$ is more metal-rich than the field at $l=+6\degr$ (K-S test p-value = 0.008) and the field at $(l=0\degr,b=1\degr)$ is more metal-poor than fields at $\vert l \vert = 6\degr$ (K-S tests p-value $<$ $7\,10^{-4}$). 


Observing further from the plane at $b=-4\degr$, \cite{Rangwala09b} find lower metallicities than we do for $l=-5.0\degr$ ([Fe/H]=$-0.17$) and $l=5.5\degr$ ([Fe/H]=$-0.55$) and a higher metallicity at $l=1\degr$ ([Fe/H]=$-0.09$), although with large individual measurement errors.
\cite{Rangwala09b} conclude that they found indications of a metallicity gradient with Galactic longitude, with greater metallicity in Baade's window. Based on the same longitudes but at $\vert b \vert \leq 1\degr$, we also find what looks like a major axis gradient but in the opposite direction, the most metal-poor population being at the centre. 

Looking at the minor axis, we find a significantly different distribution in our red clump stars at $(l=0\degr,b=1\degr)$ compared to Baade's Window values of \cite{Hill11}. On their red clump sample \cite{Hill11} find a mean [Fe/H]=0.05$\pm$0.03 with a dispersion of 0.41$\pm$0.02~dex, and the comparison with our low-latitude field leads to a significant Kolmogorov-Smirnov test p-value of 0.001. Our sample is therefore more metal-poor than at $b=-4\degr$. 
We note, however, that our metallicities are derived with a different method, that our errors are large, and that our sample target selection box is done in the near-infrared, while it is done in the optical in \cite{Hill11}. But both samples were aiming to observe the bulk of stars, and we checked in Sect. \ref{sec:targetseltest} that the only bias expected from our target selection on the metallicity distribution are against the most metal-poor stars. 

Observing 110 inner M-giants, \cite{Ramirez00} found no evidence of a metallicity gradient along the minor or major axes of the inner bulge ($R<560$~pc) and derived a mean value of [Fe/H] = $-0.21$ dex with a dispersion of 0.30 dex. 
\cite{Rich07inner} observed 17 M-giants of at $(l=0\degr,b=-1\degr)$ and found [Fe/H] = $-0.22\pm0.03$ with a dispersion of 0.14$\pm$0.02 dex. 
The mean value of both \cite{Ramirez00} and \cite{Rich07inner} is in excellent agreement with our sample at $(l=0\degr,b=1\degr)$. \cite{Rich07inner} and  \cite{Ramirez00} find no significant difference with their M-giant sample in Baade's window, although both distributions show a small metal-poor peak in their inner sample (Fig. 11 of \citealt{Ramirez00}; Fig. 2 of \citealt{Rich07inner}).

It therefore seems from our sample and its comparison with other observations that the metallicity gradient observed at $\vert b \vert > 4 \degr$ along the minor axis by \cite{Zoccali08} is also found in the plane at $l=\pm6\degr$. At $l=0\degr$, the metallicity gradient of the minor axis flattens, with a hint of an inversion, when comparing the Baade's window sample with our sample at $(l=0\degr,b=1\degr)$. However, considering the large errors of our sample and the difference in target selections and spectral analysis with other samples, a large homogeneous survey is needed to confirm this inversion. 

Despite the quoted uncertainties, we can confirm the flattening of the metallicity gradient along the minor axis within $\vert b \vert < 4 \degr$. Indeed when adding a systematic of 0.2~dex on our mean metallicity estimate, which is an upper limit according to our sensitivity analysis in Sect. \ref{Smetest}, our sample at $(l=0\degr,b=1\degr)$ becomes consistent with the mean metallicity of Baade's Window. Our results show that the metal-poor population observed at $b=-4\degr$ is at least as large at $b=1\degr$. This metal-poor population therefore does not just appear in Baade's window owing to a fading of the metal-rich one, it has a significant contribution at all latitudes including in the inner regions.

\begin{table*} 
\caption{SEMMUL Gaussian-component decomposition on the metallicity distribution. $\langle \Vgc \rangle$ is the mean galactocentric radial velocity and $\sigma_{r}$ the radial velocity dispersion. The distance to the Sun corresponds to the mode of the distance distribution function.  $\sigma_\mathrm{dist}$ corresponds to the 68\% HDI of the distribution (see section \ref{Sdistres}). The errors are computed using bootstrap resampling.}
\label{tab:Semmul}
\begin{center}
\begin{tabular}{llllllll}
\hline\hline
Field & [Fe/H] & $\sigma_{\mathrm{[Fe/H]}}$ & \% & $\langle \Vgc \rangle$ & $\sigma_{\Vr}$ & dist & $\sigma_\mathrm{dist}$ \\ 
 & dex & dex & & $\kms$ & $\kms$ & kpc & kpc \\
\hline
$l=+10\degr$ & -0.17 $\pm$ 0.13 & 0.45 $\pm$ 0.06 & 21 $\pm$ 8 & 46 $\pm$ 29 & 90 $\pm$ 16 & 4.2 $\pm$ 0.2 & 1.8 $\pm$ 0.1\\
  & 0.40 $\pm$ 0.03 & 0.16 $\pm$ 0.02 & 79 $\pm$ 8 & 57 $\pm$ 10 & 85 $\pm$ 6 & 3.9 $\pm$ 0.1 & 1.8 $\pm$ 0.1 \\
\hline
$l=+6\degr$ & -0.32 $\pm$ 0.06 & 0.39 $\pm$ 0.08 & 27 $\pm$ 4 & 31 $\pm$ 23 & 99 $\pm$ 19 & 5.7 $\pm$ 0.2 & 2.2 $\pm$ 0.1 \\
 & 0.32 $\pm$ 0.02 & 0.17 $\pm$ 0.02 & 73 $\pm$ 4 & 53 $\pm$ 19 & 129 $\pm$ 7 & 5.8 $\pm$ 0.1 & 1.9 $\pm$ 0.1 \\
\hline
$l=0\degr$, b=1$\degr$ & -0.79 $\pm$ 0.09 & 0.31 $\pm$ 0.05 & 38 $\pm$ 7 & 10 $\pm$ 17 & 134 $\pm$ 13 & 7.2 $\pm$ 0.1 & 2.7 $\pm$ 0.1 \\
 & 0.15 $\pm$ 0.05 & 0.27 $\pm$ 0.04 & 62 $\pm$ 7 & -7 $\pm$ 20 & 152 $\pm$ 8 & 7.4 $\pm$ 0.1 & 2.2 $\pm$ 0.1 \\
\hline
$l=-6\degr$ & -1.04 $\pm$ 0.10 & 0.29 $\pm$ 0.07 & 16 $\pm$ 3 & -43 $\pm$ 25 & 85 $\pm$ 19 & 8.8 $\pm$ 0.2 & 3.7 $\pm$ 0.2 \\
 & 0.23 $\pm$ 0.03 & 0.30 $\pm$ 0.02 & 84 $\pm$ 3 & -86 $\pm$ 12& 112 $\pm$ 4 & 9.2 $\pm$ 0.1 & 2.4 $\pm$ 0.1 \\
\hline
\end{tabular}
\end{center}
\end{table*}

\subsection{\label{Sdistres}Distances distribution}

\begin{figure}
 \centering
 \includegraphics[width=9cm]{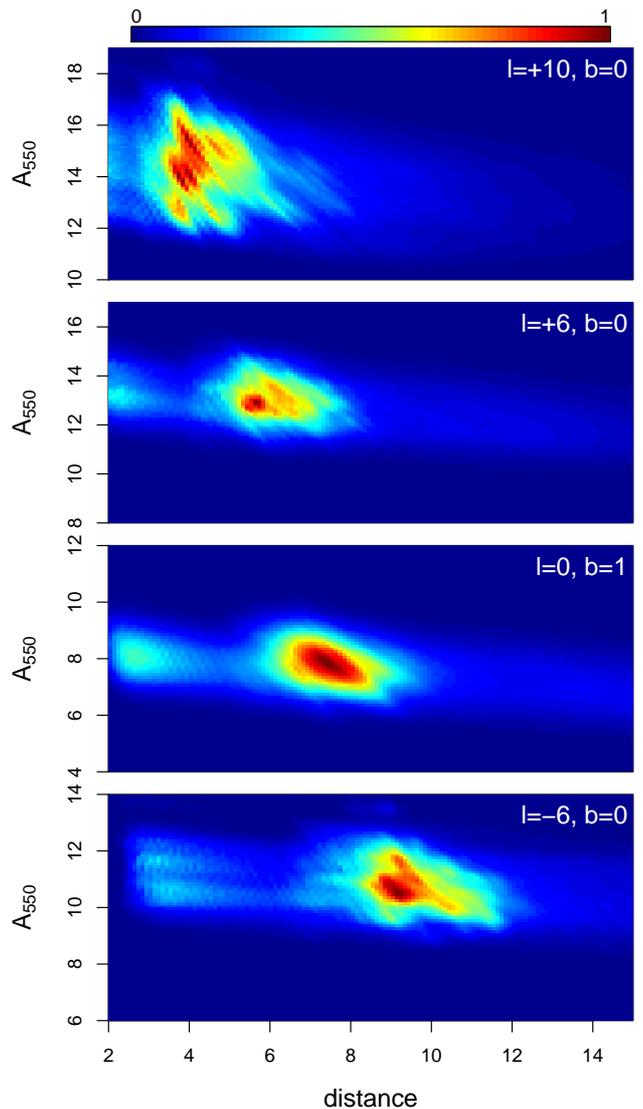}
 \caption{Density plot of the (distance, extinction) distribution of our stars. The distance is from the Sun in kpc. $A_{550}$ is the absorption at 550~nm (see Section \ref{Sdistest}). The number of stars per unit area is normalized to a value of 1.0 at the maximum density (separate normalization in each plot).}
 \label{fig:AvDistMat}
\end{figure}

The sum of the probability distribution function in the absorption-versus-distance plane of all the stars of the same fields is shown in Fig. \ref{fig:AvDistMat}. We see there the distance change along the major axis following the bar angle. 
This representation allows the main trends, the dispersion around those due to the uncertainty in each individual estimate, and the correlations between distance and extinction intrinsic to the method to be seen at the same time. The impact of the distance / extinction correlation due to our method is a spread going in the opposite direction to the real behaviour: the extinction decreases with increasing distance. The density peak at $\sim3$~kpc is intrinsic to the method since foreground stars have a higher IMF weight, so they introduce a solution with a higher extinction and shorter distance. This foreground solution could have been removed by introducing the cone effect, but we did not want to add any prior information on the parameter we wanted to derive. 

Table \ref{tab:dist} shows the mode of the distance distribution of our samples and its dispersion. The dispersion is defined here as the 68\% highest Bayesian confidence interval (or highest density interval, hereafter HDI). It includes both the intrinsic distance spread of the sample and the individual distance degeneracies. In all our fields, the distance distributions are skewed toward larger distances than the mode. 

Table \ref{tab:dist} also compares the distances of our spectroscopic samples with the red clump peak distance derived from the photometry of the full sample by \cite{Babusiaux05}. 
Here we do not rescale the distance modulus toward having the Galactic centre at 8~kpc as done in the final table of \cite{Babusiaux05} to ensure consistency with the distances derived in our spectroscopic sample. Indeed we used our simulation presented in Sect. \ref{sec:targetseltest} to check that our isochrone priors are equivalent to assuming $M_K = -1.61$~mag for the red clump magnitude, which corresponds to the solar neighbourhood value \citep{Alves00}. Keeping this distance scale also allows comparison with the studies of \cite{Nishiyama05} and \cite{Gonzalez11bar}. We keep in mind that a small population correction most probably should be applied \citep{SalarisGirardi02}, which would make the distance modulus of $(l=0\degr,b=1\degr)$ consistent with being at 8~kpc \citep{Babusiaux05}. This range of population correction is compatible with the latest estimates of the distance to the Galactic centre \citep{Eisenhauer05,Ghez08,Schonrich12}.

\begin{figure}
 \centering
 \includegraphics[width=9cm]{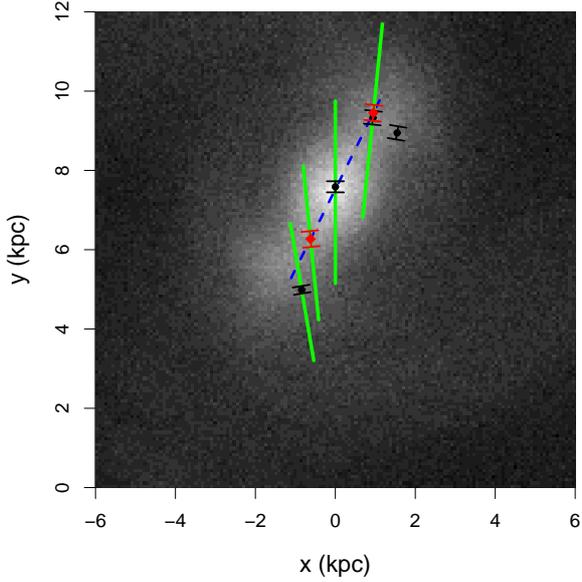}
 \caption{Line of sight and distances probed projected on the Galactic plane as seen from the north Galactic pole. The background is the \cite{Fux99} disc/bar particles within 500 pc from the Galactic plane translated to $R_0=7.6$~kpc and with an angle $\phi_{bar}=26\degr$ relative to the Sun-galactocentric line of sight. To guide the eye, the dotted line represents a straight bar of radius 2.5~kpc and with an angle $\phi_{bar}=26\degr$. The green full lines correspond to our sample locations (corresponding to 68\% HDI of the distance distribution of each sample). The black circles correspond to the density peak of \cite{Babusiaux05}. The red diamonds indicate the distance of the change in mean radial velocity used here to constrain the bar angle.}
 \label{fig:planeproj}
\end{figure}

Figure \ref{fig:planeproj} shows the projection of those distances on the Galactic plane as seen from the north Galactic pole. In this plot straight lines present the distance spread of our stars, defined as corresponding to the 68\% HDI of the sample distance distribution. To guide the eye, we indicate in this figure the location of a straight bar of radius 2.5~kpc and with an angle of 26$\degr$ relative to the Sun-Galactic centre line of sight, as well as the stellar disc particles distribution of the \cite{Fux99} model rotated also to 26$\degr$ and centered at 7.6~kpc.

Our distance is always closer than the one derived by \cite{Babusiaux05}, as expected by our target selection (Sect. \ref{sec:target}). The difference between the mean distance of our sample and the one derived by \cite{Babusiaux05} at $l=-6\degr$ and $l=0\degr$ corresponds to a difference in distance modulus of less than 0.1 magnitude. At $l=+6\degr$ we are within the first structure detected in \cite{Babusiaux05}, who detected a second structure in this field at $\sim11$~kpc, too far away to be observed in this sample. 

For the field at $l=10\degr$, the difference between our target selection distance and the over-density peak is the greatest, corresponding to 0.5 magnitude, indicating that our spectroscopic sample is significantly biased toward closer stars. This field is the only one showing a strong increase in extinction with distance in Fig. \ref{fig:AvDistMat}. This indicates that we are crossing the inner disc in this field. We know that there is a hole both in stars and in dust within the bar radius (see e.g. \citealt{Marshall06}). This hole separates nicely the bulge/bar area from the inner disc in the other fields, but it seems that it does not separate it that well at $l=10\degr$. At $l=10\degr$ \cite{Babusiaux05} detected both a high-extinction cloud in the disc at $\sim$3.5~kpc and the over-density linked to the bar at 5.1~kpc, and our target selection seems to fall between those two features. Still, about a quarter of the sample should be within the near side of the main over-density according to our distance distribution. 

The over-density seen at $l=10\degr$ in the near-infrared colour-magnitude diagram is also seen with 6.7 GHz methanol masers by \cite{Green11} within $10\degr<l<14\degr$ and would therefore correspond to a young structure, consistent with the metal-rich distribution of our sample. According to the structure of the bar traced by red clump stars (Fig. 3 of \citealt{Gonzalez11bar}), it indeed seems most likely that all the components (end of the inner disc, molecular/stellar pseudo-ring and bar) start to merge at this longitude.  

The spread of the distances sampled in this study is particularly large at $l=-6\degr$. According to the  Gaussian decomposition on the metallicity distribution presented in Table \ref{tab:Semmul}, the component with a mean metallicity of [Fe/H]=$-1$~dex is located at a mean distance of 8.7~kpc, significantly closer to us than the metal-rich component. It therefore seems that our line of sight first crosses a region dominated by the metal-poor component before reaching the bar.


\begin{table}
\caption{Mode and dispersion (defined as the 68\% HDI) of the distance distribution in our sample, in kpc from the Sun. $D_{phot05}$ corresponds to the field distance derived by \cite{Babusiaux05} before scaling the field $l=0\degr,b=1\degr$ to 8~kpc.
$A_{550}$ and  $\sigma_{A}$ also provide the mode and dispersion of the extinction distribution of the samples, $A_{550}$ being the absorption at 550~nm (see Section \ref{Sdistest}).}
\label{tab:dist}
\begin{tabular}{llllll}
\hline\hline
Field & distance &  $\sigma_{d}$ & $D_{phot05}$ & $A_{550}$ & $\sigma_{A}$ \\
 & kpc & kpc & kpc & mag & mag \\
\hline
$l=+10\degr, b=0\degr$ & 4.0 & 1.8 & 5.1 & 14.0 & 1.6 \\ 
$l=+6\degr, b=0\degr$ & 5.7 & 2.0 & $\sim 6$ & 12.9 & 1.0 \\ 
$l=0\degr, b=1\degr$ & 7.3 & 2.3 & 7.6 & 7.6 & 0.8 \\ 
$l=-6\degr, b=0\degr$ & 9.2 & 2.5 & 9.4 & 10.6 & 1.0 \\ 
\hline
\end{tabular}
\end{table}

Table \ref{tab:dist} also provides the mode of the extinction distribution and its dispersion for our samples. \cite{Omont99} derived for our field at $l=0\degr$ a fairly uniform extinction of $A_V=5.8\pm1$~mag by fitting theoretical isochrones to the red-giant branch, in agreement with Fig. 4 of \cite{Babusiaux05}, but much less than our derived value of 7.7~mag. To avoid the difference in the extinction law treated more rigorously here, we translated those values in $E(J-K)=0.97$ for the \cite{Omont99} study against $E(J-K)=1.17$ here. A 0.2~mag remaining difference could be explained by the difference in the isochrone set and in particular by the fact that both \cite{Omont99} and \cite{Babusiaux05} use a solar metallicity isochrone as a reference, while one expects the red clump colour to decrease with decreasing metallicity. This would confirm that the low S/N in our spectra is indeed due to an under-estimated visual extinction during the target selection.

\cite{Gonzalez12ext} have derived the mean extinction of the bulk of red clump stars in the VVV fields by comparison with Baade's window red clump colours. They derived $E(J-K)=(2.46\pm0.49, 2.28\pm0.45, 1.15\pm0.14, 1.80\pm0.24)$ for our fields at $l=(10\degr, 6\degr, 0\degr, -6\degr)$, to be compared to our samples mode $E(J-K)=(2.10, 1.93, 1.16, 1.59)$. As expected  from our target selection, we always have a slightly lower extinction value than the bulk of stars in those fields, although within the spread. Only the field at $l=0\degr$ has a mean extinction fully consistent with the \cite{Gonzalez12ext} map, which is due to the low extinction spread in this field.

\subsection{\label{Svrdist}Radial velocities versus distance}

\begin{figure}
 \centering
 \includegraphics[width=8cm]{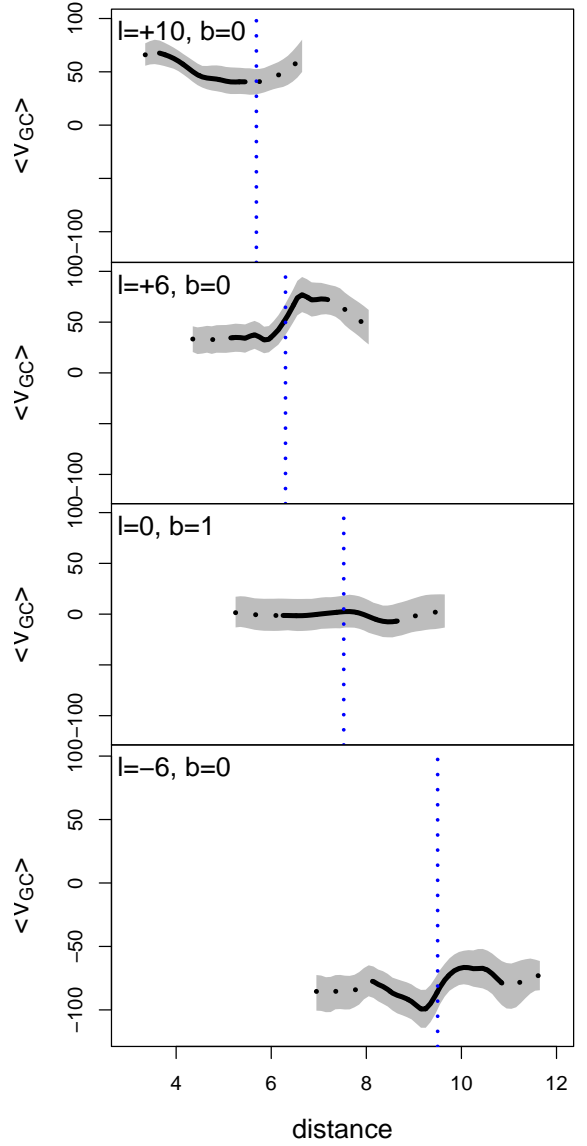}
 \caption{Mean Galactocentric radial velocity as a function of distance from the Sun. Only the distances corresponding to the full sample highest Bayesian confidence interval (HDI) are considered. 
 Dotted black lines correspond to the 68\% HDI and full lines corresponds to the 50\% HDI. 
 In grey the confidence interval at 1 sigma obtained by bootstrap.
 The vertical dotted blue lines indicate the distance corresponding to the centre of a bar with $R_0=7.6$~kpc and $\phi_{bar}=26\degr$, as represented in Fig. \ref{fig:planeproj}. 
 }
 \label{fig:distVr}
\end{figure}

Figure \ref{fig:distVr} shows the mean Galactocentric radial velocity as a function of distance. For this we computed the mean of each stellar velocity weighted by its distance probability on a fine distance grid. We only plotted the distance range were the bulk of stars are located, to ensure that the shape of the result is not distorted by only a few stars' probability tails. We defined this distance range by the highest Bayesian confidence interval on the full-sample distance distribution.  
Figure \ref{fig:distVr} shows the distances corresponding to both the 50\% and 68\% HDI. At the border of the distance distribution the number of stars contributing to the mean value decreases leading to the tendency of flattening the mean velocity versus distance relation. More generally, the large spread of our distance distribution function also leads to a flattening of the curves. Since any real velocity versus distance relation will be flattened by our distance errors, any remaining variation observed in our plots is likely to be a strong feature. 

The field at $l=-6\degr$ shows  a small variation in the radial velocity with distance, indicating a minimum value occurring at 9.2$\pm0.2$~kpc, just before the distance of the peak in density (Table \ref{tab:dist}). It corresponds to the high-velocity component described in Table \ref{tab:SemmulVr} smoothed by the large spread of each individual star distance distribution.

The field at $l=+6\degr$ shows a significant variation in the radial velocity with distance. It was also detected in our sample prior to individual distance probability determination using the reddening-independent magnitude $K_{J-K}$ as a distance probe: by separating the sample along the median $K_{J-K}$ we find that the bright stars and faint stars have a significantly different mean radial velocity (Welch Two sample t-test p-value=$3\,10^{-4}$). The change in the mean velocity is located at 6.3$\pm0.2$~kpc, just after the mode of the distance distribution of our full sample and before the position of the high-velocity component described in Table \ref{tab:SemmulVr}. 

The velocity variation with distance was also detected at $l=\pm5\degr$ at $b=-3.5\degr$ by \cite{Rangwala09a}. Their figure 12 and our Fig. \ref{fig:distVr} are fully compatible: at $l=-6\degr$ they see a minimum just before their peak in magnitude, while at $l=+6\degr$ they see a sharp increase in velocity around their peak in magnitude. 

\begin{figure}
 \centering
 \includegraphics[width=8cm]{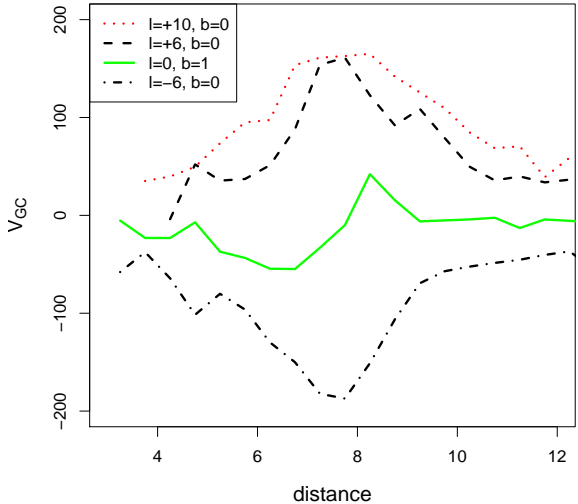} 
 \caption{Streaming motion of the stars as predicted by the \cite{Fux99} model (bar/disc particles only) in our lines of sight. Angle (26$\degr$) and distance to the Galactic centre (7.6~kpc) are the same as in Fig. \ref{fig:planeproj}. Typical bootstrap errors are $\sim 10~\kms$.}
 \label{fig:distVrFux}
\end{figure}

A solid-body cylindrical rotation would show no variation in the radial velocity with distance. A variation in radial velocity as a function of distance is instead expected for stars streaming along the near and far side of the bar (e.g. \citealt{Hafner00}, \citealt{MaoPaczynski02} and Fig. \ref{fig:distVrFux}). 
A bar streaming motion was also detected in the plane by SiO masers \citep{Deguchi00b}. It was detected along the minor axis in the radial velocities of the metal-rich population of Baade's window $(l=1\degr,b=-4\degr)$ by \cite{Babusiaux10} as well as in the proper motions by \cite{Sumi03}, in the radial velocities at $(l=0\degr,b=-5\degr)$ by \cite{Ness12} and $(l=0\degr,b=-6\degr)$ by \cite{Vasquez13}, in the HST proper-motion at $(l=1\degr,b=-3\degr)$ by \cite{Clarkson08}, and in the OGLE-III proper motions by \cite{Poleski13}. However $l=0\degr$ is our only field without any sign of variation of the radial velocity with distance. Selecting only metal-rich stars does not change this, unlike Baade's window results, although the sample here is much smaller. 

The field at $l=10\degr$ shows a small decrease in radial velocity with distance at distances smaller than $\sim$5~kpc. Close-by stars have a mean velocity of $\langle \Vgc \rangle =68\pm11~\kms$, consistent with the BRAVA and APOGEE results (Fig. \ref{fig:VrBrava}). We have seen that the closest stars are in fact inner disc stars. At the main CMD over-density (5.1~kpc), the radial velocity is $\langle \Vgc \rangle =42\pm12~\kms$. 
There is a hint that the mean velocity goes up again at larger distances. There are few stars at those distances, but we would have expected the distribution at the borders to be flattened by the spread of the distance distribution, while a signal is still here, which means that it is quite likely that this feature is real. 

Figure \ref{fig:distVrFux} presents the radial velocity variation with distance predicted by the model of \cite{Fux99} in our lines of sight. 
The predicted trends correspond very well to the observations at $l=+6\degr$. As expected, the amplitude of the variation as predicted by the model is seen flattened in our data by the spread in our distance distributions. At $l=10\degr$ and distances lower than 5~kpc, there are very few particles in the model. At distances greater than 5~kpc, it is our sample that is small but the increase in radial velocity predicted by the model can be seen at the border of our distance distribution.
At $l=0\degr$ a small variation is predicted by the model but not seen here. It may simply be due to the smoothing of this small variation. 
At $l=-6\degr$, the minimum of the velocity is seen in the model at $R_0=7.6$~kpc, while we see it at $9.2\pm0.2$~kpc. Changing the angle of the bar in the model does not change the result, because at $l=-6\degr$ we are  crossing the nuclear bulge in the model \citep[see][Figs. 18 and 23]{Fux99}, which extends in the model up to $l=-7\degr$. It therefore seems that we can give $l=-6\degr$ as an upper limit for the extent of the nuclear part of the bulge, in agreement with the flattening observed in the red clump distribution within $\vert l \vert < 4\degr$ found by \cite{Nishiyama05} and \cite{Gonzalez11bar}.

Since we expect the change in radial velocity with distance observed at $l=\pm6\degr$ to be due to the bar streaming motion, we would also expect that the distance at which this change in mean radial velocity occurs is linked to the bar angle. As a first test, we used our derived distance to this break in the radial velocity behaviour (6.3$\pm0.2$~kpc at $l=+6\degr$ and 9.5$\pm0.2$~kpc at $l=-6\degr$) as a distance indicator for the bar. Those are indicated in Fig. \ref{fig:planeproj}. With those two points we derived an angle for the bar of $\phi_{bar}=26\pm3\degr$ and a distance from the Sun to the Galactic centre $R_0=7.5\pm0.2$~kpc (represented in Figs. \ref{fig:planeproj} and \ref{fig:distVr}). The quoted errors are issued from bootstrap alone and do not include any systematics. All this seems self-consistent but needs to be confirmed by studies at other longitudes. That the distance derived by the mean radial velocity change corresponds to the distance of the peak of the red clump in the CMD could seem at odds with the bias of the latter due to the bar thickness predicted by \cite{Stanek94} and \cite{GerhardMartinezValpuesta12}. We also tested this with the model of \cite{Fux99} with the same conclusions, but by inspecting the simulated CMDs at $l=-6$,$+6$ and $+10\degr$, the red clump peak is much more diluted in the simulation than in the observations of \cite{Babusiaux05}. While \cite{Gonzalez11bar} observed at $\vert b \vert = 1\degr$, leading to a bar shape that is reproduced well by the model of \cite{GerhardMartinezValpuesta12}, we are observing here at $b=0\degr$, an hypothesis could therefore be that we are looking at younger stars associated with shocks along the bar, creating a sharper red clump than predicted in the N-body simulations that do not include star formation.

\subsection{Radial velocities versus metallicity}

\begin{figure*}
 \centering
 \includegraphics[width=18cm]{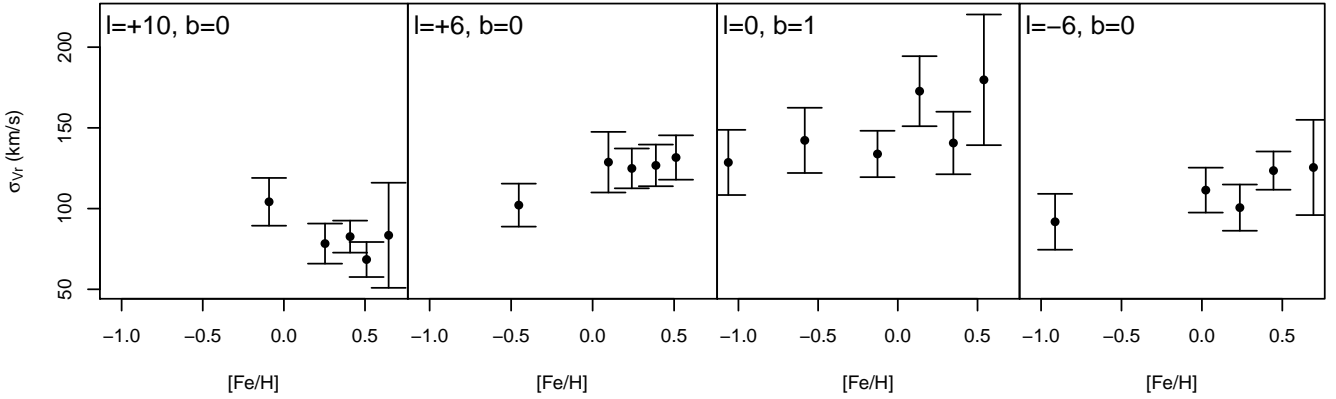}
 \caption{Radial velocity dispersion as a function of metallicity by bins of 20 stars. Error bars are obtained by bootstrap.}
 \label{fig:metdispVr}
\end{figure*}

\begin{figure*}
 \centering
 \includegraphics[width=8cm]{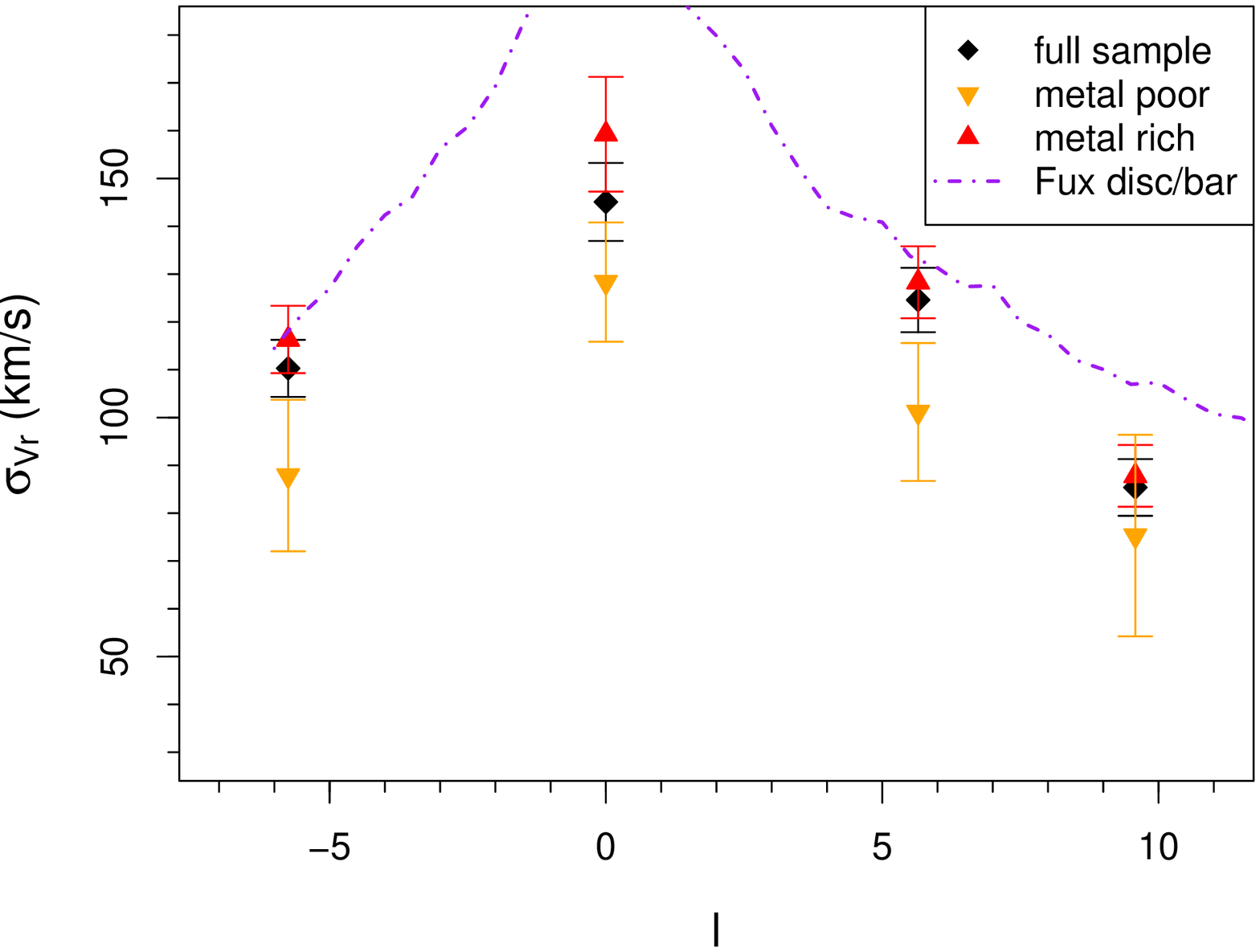}
 \hspace{1cm}
 \includegraphics[width=8cm]{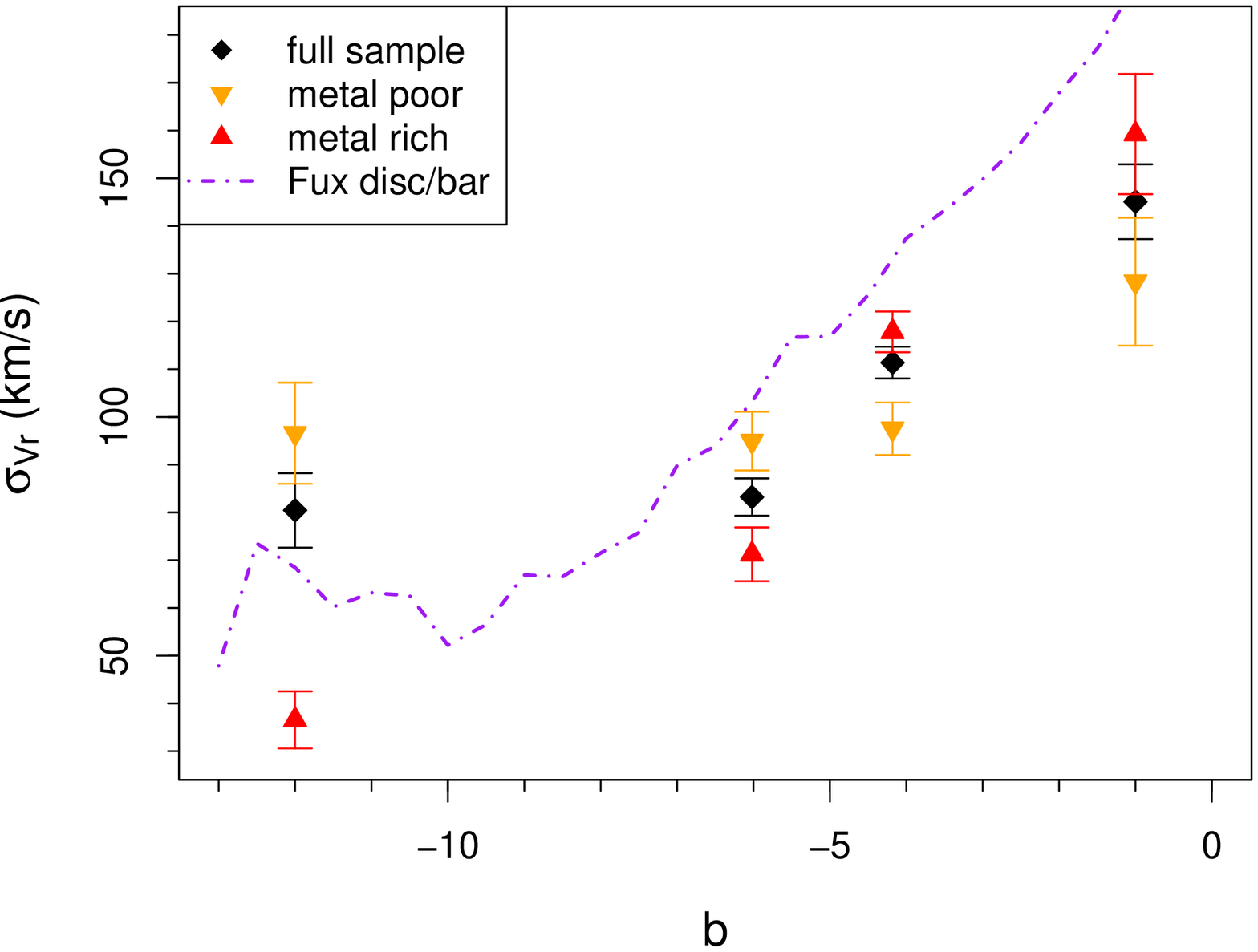}
 \caption{Radial velocity dispersion along the bulge's a) major and b) minor axes compared with the disc particles of the model of \cite{Fux99} rotated and translated as in Fig. \ref{fig:planeproj}. 
The metal-poor population corresponds to [Fe/H]$<-0.2$~dex and the metal-rich population to [Fe/H]$>$0~dex.
Figure b) completes Fig. 8 of \cite{Babusiaux10} with our minor axis field $(l=0\degr,b=1\degr)$ assuming symmetry between positive and negative latitudes.
}
 \label{fig:compFux}
\end{figure*}

Figure \ref{fig:metdispVr} shows the radial velocity dispersion as a function of metallicity. 
In all fields but $l=10\degr$, the velocity dispersion is higher for the metal-rich stars than for the metal-poor stars. The highest velocity dispersion is seen for the metal-rich stars at $l=0\degr$. 
Figure \ref{fig:compFux} shows the velocity dispersion as a function of longitude and latitude, the latter using the data of \cite{Babusiaux10} completed by our $l=0\degr$ field. As for Fig. 8 of \cite{Babusiaux10}, we separated the sample into a metal-poor and a metal-rich population, although the cut is done here with fixed metallicity ([Fe/H]$<-0.2$ for the metal-poor, [Fe/H]$>0$ for the metal-rich). The \cite{Fux99} model has been rotated and translated as indicated in Fig. \ref{fig:planeproj}, which corrected for a shift in the barycentre position of the model, leading to an increase in the velocity dispersion for the disc/bar particles along the minor axis compared to the original model used in \cite{Babusiaux10}. Along the major axis, the particules have been selected to be within 2~kpc of the bar mean position assuming $\phi_{bar}=26\degr$.
We did not add the spheroidal component of the \cite{Fux99} model in this plot since in \cite{Babusiaux10} we have shown that this component had a velocity dispersion that is too large, even if it predicted that the velocity dispersion of this component stayed fairly constant with Galactic latitude, as observed. We see here that the velocity dispersion is also higher than the observations for the disc/bar particles. 
Although the error bars are large, it continues the trend observed in \cite{Babusiaux10}: the velocity dispersion of the metal-rich stars increases going near the Galactic plane, following the trend of the \cite{Fux99} disc/bar particles. 
The metal-poor star velocity dispersions seem to stay around 100 $\kms$ along the major axis as it did along the minor axis. The highest dispersion of the metal-poor stars is seen at $l=0\degr$, which could be interpreted as the influence of the bar dynamics on an older metal-poor structure \citep{Saha12}. 

We tested a 2D Gaussian-component decomposition using metallicity and radial velocity as discriminating variables. The result is a combination of Table \ref{tab:SemmulVr} (using only $\Vr$) and \ref{tab:Semmul} (using only [Fe/H]): the metal-poor component in all fields stays as indicated when using only metallicity as a discriminant. In fields l=$\pm6\degr$ the metal-rich component ([Fe/H]=0.2~dex) splits in a component with a low mean galactocentric radial velocity and a high-velocity component ($\vert \Vgc \vert=200\pm15\kms$), confirming that this high-velocity component corresponds to the tail of the bar's velocity distribution. 

The field at $l=+10\degr$ shows the lowest velocity dispersion in the metal-rich regime. We have seen that $l=+10\degr$ actually corresponds to the end of the inner disc and the molecular/stellar ring where we indeed would expect a lower radial velocity dispersion than within the bar. High metallicity in this region that is rich in gas is also expected. In Table \ref{tab:Semmul} there is a hint that the metal-poor component in this field is located a bit further from the metal-rich component. To test this, we tried a 2D Gaussian decomposition using both the metallicity and the distance as discriminating variables. It leads to three components which confirms this hypothesis: 56\% of the sample is in a cluster with [Fe/H]=0.4~dex, a distance of 3.9$\pm$0.15~kpc and a velocity dispersion of 74$\pm11~\kms$, which we would associate to the end of the inner disc. 39\% of the sample is in a cluster with [Fe/H]=0.22~dex, a distance of 5.2$\pm$0.41~kpc and a velocity dispersion of 107$\pm15~\kms$, which we would associate to the start of the bar. The remaining 6\% are the most metal-poor stars with a distance in between of 4.5$\pm$0.17~kpc.

\section{\label{Sconclu}Conclusions}

We analysed the low-resolution spectra of red clump stars in four fields along the bar's major axis.
The main results of our analysis can be summarized as follows: 
\begin{itemize}
\item  We observed an increase in the radial velocity dispersion of the bar near the galactic plane compared to literature values further from the plane. 

\item  Our field at $l=+10\degr$ seem to sample both the end of the inner disc and the bar. The end of the inner disc is the largest component in our sample. It shows a large spread in both distance and extinction, a high metallicity and a low mean velocity and velocity dispersion. Our target selection shows that the bar is no longer as distinct from the inner disc as it is in the other fields and that we are therefore sampling here a field close to the end of the bar.  

\item  We detected the streaming motion of the stars induced by the bar in fields at $l=\pm6\degr$. 

\item We confirmed that the high-velocity component detected by \cite{Nidever12} is associated with the bar streaming motion and derive a mean metallicity of $0.23\pm0.07$~dex for this component. 

\item From the distance at which the radial velocity shows a sharp change in its mean value, we tentatively estimated a bar angle of $\phi_{bar}=26\pm3\degr$. (The quoted error does not include systematics due to the choice of priors in the distance estimations.)


\item  All our fields show a significant fraction of metal-poor stars ([Fe/H]$<-0.5$), the largest population being at $l=0\degr$. At $l=-6\degr$ the large spread in distance seems to be due to the fact that we are crossing such a metal-poor component before reaching the bar. This could indicate that the metal-poor population is spread more uniformly within the inner region than the metal-rich population, which is concentrated along the bar.  

\item  We confirmed that the metallicity gradient observed at $\vert b \vert>4\degr$ flattens in the inner regions, with a hint that the gradient is actually inverted. The metal-poor population therefore does not appear only at high latitudes owing to a fading of the metal-rich one. It has a significant contribution at all latitudes including in the inner regions. 

\end{itemize}

All those results are consistent with the expected kinematic signature of a secular bar, as predicted by dynamical models of the bar and with the bulge being composed of both a metal-poor and a metal-rich component with different density distributions. 
What we call here the metal-rich component would correspond to the populations A \& B of \cite{Ness13kine} and the bar population of \cite{Robin12} ($\mathrm{[Fe/H]}\gtrsim-0.5$). The metal-poor component would correspond to the populations C \& D of \cite{Ness13kine} and the ``thick bulge'' population of \cite{Robin12} ($\mathrm{[Fe/H]}\lesssim-0.5$). The metal-rich population is the population that presents a vertex deviation in Baade's window \citep{Babusiaux10} and which follows the X-shape structure in \cite{Ness12}. 
It issues from the disc through secular evolution.
The metal-poor population is more centrally concentrated, extends further from the Galactic plane, and presents a kinematically distinct signature. 

Our data are consistent with a main bar with an angle of $\phi_{bar}=26\pm3\degr$ and length of $\sim$2.5~kpc, containing a nuclear bulge in its centre extending not further than $\sim$560~pc. Such a size for the nuclear bulge would be consistent along the major axis with the fact that we are not crossing this structure at $l=-6\degr$ and with the flattening of the bar angle at $\vert l \vert<4\degr$ detected by \cite{Nishiyama05} and \cite{Gonzalez11bar}, and along the minor axis with the flattening of the metallicity gradient observed at $\vert b \vert<4\degr$. What we call here a nuclear bulge is not necessarily a distinct dynamical structure from the main bar but could be due to the bar inner Lindblad resonance (ILR) \citep{Fux99,GerhardMartinezValpuesta12}. This dynamical structure could mix the disc and a more primordial structure trapped within the ILR.  

The central concentration of metal-poor stars is also observed in the RRLyrae \citep{Alcock98,Collinge06,Pietrukowicz12} and Type II Cepheids \citep[][Fig. 6]{Soszynski11_cepheids} distributions. \cite{Collinge06} and \cite{Pietrukowicz12} show that the RRLyrae follow the barred distribution of the bulge red clump giants in the inner regions ($\vert l\vert < 3\degr , \vert b \vert < 4\degr$), while farther off the
Galactic plane ($\vert b \vert > 4\degr$), the distribution of RRLyrae stars become spherical. This result would be consistent with the results of \cite{Saha12} who show that a low-mass classical bulge could develop triaxiality and cylindrical rotation under the influence of the bar.  A metal-poor stars concentration would also be consistent with N-body simulations showing that stars located within the central regions before the bar instability tend to stay confined in the innermost regions of the boxy bulge \citep[e.g.][]{DiMatteo13}, independently of their discy or spheroidal origin. 
We note that another interpretation could have been an inflow of metal-poor gas towards the centre through the bar, leading to younger stars forming in less enriched gas.
This inflow of gas should have occurred more than 10 Gyrs ago for this scenario to be consistent with the fact that bulge metal-poor stars have been shown to be older than 10~Gyr \citep{Zoccali03,Clarkson08,Bensby13}.

Several models study this double composition in detail \citep{Samland03,TsujimotoBekki12,Grieco12,Perez13}, where the metal-rich population is associated with the bar and the metal-poor population is either the thick disc or a primordial structure formed either by hierarchical formation, dissipational collapse, or clumpy primordial formation.
That those two (or more) populations can present internal metallicity gradients complicates the interpretation of observations even more at different longitudes and latitudes (see e.g. \citealt{Grieco12} discussing a metallicity gradient within a classical gravitational gas collapse component and \citealt{MartinezValpuestaGerhard13} discussing a metallicity gradient in a secular bulge). 
The exact density distribution of the metal-poor population and its link in terms of formation history with the thick disc and the inner halo, which are seen in their abundances and kinematics, need large homogeneous surveys. 




\begin{acknowledgements}
We would like to acknowledge Mike Irwin for providing his GIRAFFE sky-subtraction program and Roger Fux for providing his N-body model. 
We thank the referee, David Nataf, for useful comments that helped us improve the clarity of this article. 
We acknowledge the support of the French Agence Nationale de la Recherche under contract ANR-2010-BLAN-0508-01OTP. 
\end{acknowledgements}

\bibliographystyle{aa} 
\bibliography{BarGiraffe}

\end{document}